\documentclass[floatfix,showpacs,amsmath,amsfonts,amssymb,aps,twocolumn, prb,groupedaddress]{revtex4-1}

\usepackage[table,x11names]{xcolor}
\usepackage{graphicx}
\usepackage{dcolumn}
\usepackage{amsthm}
\usepackage{bm}
\usepackage{siunitx}
\usepackage{nicefrac}
\usepackage{todonotes}
\usepackage{mathtools}
\usepackage{float}
\usepackage{braket}
\usepackage{makecell} 
\usepackage{multirow}
\usepackage{soul}

\definecolor{darkred}{rgb}{0.6,0.,0.}
\definecolor{darkgreen}{rgb}{0.,0.5,0.}
\definecolor{darkblue}{rgb}{0.,0.,0.6}
\usepackage[ breaklinks, colorlinks=true, linkcolor=darkred, citecolor=darkgreen, urlcolor=darkblue]{hyperref}

\newcommand{\includegraphicsgood}{\includegraphics}

\usepackage[inactive,tightpage]{preview}
\PreviewMacro[*[[!]{\input{}}
\PreviewMacro[*[[!]{\includegraphicsgood}

\begin{document}

\title{Absence of diagonal force constants in cubic Coulomb crystals}

\author{Bartholomew~Andrews}
\affiliation{TCM Group, Cavendish Laboratory, University of Cambridge, Cambridge CB3 0HE, United Kingdom}
\affiliation{Department of Physics, University of Zurich, Winterthurerstrasse 190, 8057 Zurich, Switzerland}
\author{Gareth~Conduit}
\affiliation{TCM Group, Cavendish Laboratory, University of Cambridge, Cambridge CB3 0HE, United Kingdom}
\date{\today}


\begin{abstract}
The quasi-harmonic model proposes that a crystal can be modeled as atoms connected by springs. We demonstrate how this viewpoint can be misleading: a simple application of Gauss' law shows that the ion-ion potential for a cubic Coulomb system can have no diagonal harmonic contribution and so cannot necessarily be modeled by springs. We investigate the repercussions of this observation by examining three illustrative regimes: the bare ionic, density tight-binding, and density nearly-free electron models. For the bare ionic model, we demonstrate the zero elements in the force constants matrix and explain this phenomenon as a natural consequence of Poisson's law. In the density tight-binding model, we confirm that the inclusion of localized electrons stabilizes all major crystal structures at harmonic order and we construct a phase diagram of preferred structures with respect to core and valence electron radii. In the density nearly-free electron model, we verify that the inclusion of delocalized electrons, in the form of a background jellium, is enough to counterbalance the diagonal force constants matrix from the ion-ion potential in all cases and we show that a first-order perturbation to the jellium does not have a destabilizing effect. We discuss our results in connection to Wigner crystals in condensed matter, Yukawa crystals in plasma physics, as well as the elemental solids.       
\end{abstract}

\maketitle


The classical theory of crystal stability was extensively studied by Born in the first half of the 20th Century~\cite{Born40}. This seminal work focused on deriving the Born stability criteria based on the elasticity constants, as well as determining the scope of the Cauchy-Born rule of crystal deformation~\cite{Ericksen08}. Since this time, the topic of crystal stability has been revisited from numerous perspectives~\cite{Grimvall12}: from the historic models of ionic matter by Born-Land{\'e}~\cite{Brown02}, Born-Mayer~\cite{Born32}, and Kapustinskii~\cite{Kapustinskii56}; the Hume-Rothery rules for metal alloys~\cite{Rothery35}; through to sophisticated quantum Monte Carlo simulations in current research~\cite{Driver10,Ahn18}. However, these works are based on quadratic modes, which we demonstrate can be absent from the most basic ion-ion interaction of many common crystal structures. This motivates us to revisit the stability analysis of prototypical models, which are currently of interest to the electronic structure~\cite{Grimvall12}, plasma physics~\cite{Hansen73}, and astrophysics~\cite{Chu94} communities.

In this paper, we study the force constants matrix with respect to atomic positions for infinite crystals in the bare ionic, density tight-binding, and density nearly-free electron regimes\footnote{We use the terms ``density tight-binding model'' and ``density nearly-free electron model'' to emphasize the fact that this is in reference to the electron densities.}. Having observed that the ion-ion potential for cubic crystal structures can have no diagonal harmonic contribution, we seek to answer the question of what repercussions this has on preferred crystal structure. By looking at a variety of crystal lattices, motivated by the elemental solids in the periodic table, we draw comparisons between specific structures. We stabilize the ionic crystal for all structures through the inclusion of electrons in our model, we study the stability transition, and use our framework to unify complementary models in the literature. We show that, using this simple yet overlooked observation, insight is gained into low-energy crystal structure relaxation.

We first introduce the underlying theory in Sec.~\ref{sec:theory}. We then proceed to examine the bare ionic crystal, and subsequently the density tight-binding, and density nearly-free electron regimes in Secs.~\ref{sec:icm},~\ref{sec:tbm}, and~\ref{sec:nfem}, respectively. Finally, we summarize the conclusions and implications of the results in Sec.~\ref{sec:conclusion}.


\section{Theory}
\label{sec:theory}

We consider an infinite crystal of atoms in three-dimensions and at zero-temperature.

The general Hamiltonian of the system is
\begin{equation*}
\hat{H}=\hat{T}_\text{i}+\hat{T}_\text{e}+\hat{V}_\text{i-i}+\hat{V}_\text{e-i}+\hat{V}_\text{e-e},
\end{equation*}
where $\hat{T}_\text{i}$, $\hat{T}_\text{e}$ are the ion and electron kinetic energies, and $\hat{V}_\text{i-i}$, $\hat{V}_\text{e-i}$, and $\hat{V}_\text{e-e}$, are the ion-ion, electron-ion, and electron-electron contributions to the potential energy, respectively. 

We work in the Born-Oppenheimer (BO) approximation, where the ions are assumed to be significantly more massive than the electrons and therefore move on much longer time scales. In this approximation, the complete many-body problem may be solved in two steps: first, with the BO Hamiltonian containing only the electronic degrees of freedom and ions assumed fixed in space; and second, with the ions free to move in the previously-calculated BO potential energy surface to account for the nuclear contribution to the kinetic energy. For the computation of the inter-atomic force constants in this paper, we use the BO potential energy surface, $E^\text{BO}$~\cite{Baroni01}. In the cases where we need the total energy, $E$, we then solve the nuclear problem that includes the kinetic energy of the ions.

The contribution from the electronic kinetic energy is discussed in the sections for the models. The ion-ion, electron-ion and electron-electron potential energies are given by the Coulomb interaction. 



Each unit cell of the crystal has an atom at the origin of the cell with position $\mathbf{R}_I$ (upper case). There may also be additional atoms in the unit cell with displacement vectors $\mathbf{r}_i$ (lower case) relative to $\mathbf{R}_I$. The general position of an atom at equilibrium may be written as $\mathbf{R}^0_{Ii}=\mathbf{R}_I+\mathbf{r}_i$. We consider an instantaneous small and finite displacement $\mathbf{u}_{Ii}$ of an atom in the crystal, such that the general position of an atom is given as $\mathbf{R}_{Ii}=\mathbf{R}_I+\mathbf{r}_i+\mathbf{u}_{Ii}$.

Harmonic lattice dynamics is based on a Taylor expansion of the total energy about structural equilibrium. In the BO approximation, this yields
\begin{equation*}
E^\text{BO}(\{\mathbf{R}_{Ii}\})=E^\text{BO}(\{\mathbf{R}^0_{Ii}\}) +\frac{1}{2}\sum_{Ii\alpha, j\beta} \Phi_{Ii\alpha, 0j\beta} u_{Ii\alpha} u_{0j\beta},
\end{equation*}
where $\alpha,\beta$ are Cartesian directions and the adiabatic and harmonic approximations are assumed~\cite{Baroni01}. The quantity $\Phi_{Ii\alpha, 0j\beta}$ is known as the matrix of force constants, given as
\begin{equation*}
\Phi_{Ii\alpha, 0j\beta}=\left. \frac{\partial^2 E^\text{BO}}{\partial u_{Ii\alpha} \partial u_{0j\beta}}\right|_{\mathbf{u}=\mathbf{0}},
\end{equation*}
where $J=0$ due to translational invariance. This quantifies the stability of a crystal due to the movement of particular atoms. In periodic solids, it is common to subsequently examine the mass-reduced Fourier transform of the force constants matrix, known as the dynamical matrix, given as
\begin{equation}
\label{eq:dyn_mat}
D_{i\alpha, j\beta}(\mathbf{k}) = \frac{1}{\sqrt{m_i m_j}}\sum_I \Phi_{Ii\alpha, 0j\beta} e^{-\mathrm{i} \mathbf{k}\cdot\mathbf{R}_I},
\end{equation}
where $m_i$ is the mass of particle $i$ and $\mathbf{k}$ is the linear momentum vector. The eigenvalues of the dynamical matrix are the squared frequencies, $\omega^2$. The dynamical matrix is used to compute eigenmodes and definitively quantify whether a system is stable.

In 1904 Drude proposed the paradigmatic model of a crystalline solid to be atoms connected by springs, which implies that the atoms move in a harmonic potential~\cite{Drude1904_1, Drude1904_2, Cassidy}. Here we focus on a crucial contribution to this atom-atom potential, the ion-ion interaction, and show that the ions in cubic crystals are not necessarily bound by a harmonic potential. To demonstrate this statement, we assume a one-component ionic lattice in the absence of any background charge and analyze the components of the matrix of force constants in turn.

First, we examine the diagonal matrix of force constants with respect to the motion of a single ion, $\Phi_{0i\alpha,0i\beta}$. Since the second derivative is with respect to the position of a single ion, $\Phi_{0i\alpha,0i\beta}=0$ for $\alpha\neq\beta$ by symmetry. Furthermore, the sum over Cartesian directions for this matrix of force constants is equivalent to the Laplacian of the BO energy, such that $\sum_\alpha \Phi_{0i\alpha,0i\alpha} = \nabla_{0i}^2 E^\text{BO}$. Hence, the Poisson equation of Gauss' theorem demands $\sum_\alpha \Phi_{0i\alpha,0i\alpha} = 0$, which is true for all crystal structures. For cubic crystals, symmetry implies $\Phi_{0i\alpha,0i\alpha} = 0$ and therefore the ions are not harmonically bound, whereas for non-cubic crystals, $\sum_\alpha \Phi_{0i\alpha,0i\alpha} = 0$ implies that some terms will be positive and other terms will be negative.

Second, we examine the matrix of force constants, $\Phi_{Ii\alpha,0i\beta}$, with respect to the motion of two ions that reside in different unit cells, one in cell $0$ and the other in cell $I$\footnote{$J=0$ without loss of generality.}. In this case, the second derivative is mixed and so the sum over Cartesian directions no longer corresponds to the Laplacian. After perturbing the ion in unit cell $0$, we subsequently need to perturb the corresponding ion in unit cell $I$ to obtain the cross terms. We demonstrate in Sec.~SI that for all crystal structures we obtain the result $\sum_\alpha \Phi_{Ii\alpha,0i\alpha}=0$. We note that this two-ion result with both ions moving parallel to each other has a pleasing analogy to the Poisson equation for the motion of a single ion. In Sec.~SI, we also consider the non-parallel motion of the two ions, which for centrosymmetric crystals yields the corollary $\sum_{I\neq0} \Phi_{Ii\alpha,0i\beta}=0$, where the summation is over unit cells $I$. To include the $I=0$ term in the summation, which corresponds to the motion of a single ion, we can use the result from the previous paragraph that $\Phi_{0i\alpha,0i\beta}=0$ for cubic crystals. This implies that the full sum $\sum_{I} \Phi_{Ii\alpha,0i\beta}=0$ holds for all centrosymmetric cubic crystals, which includes all of the cubic space groups considered in this paper.

Using the above analysis, we have shown that $\sum_\alpha \Phi_{Ii\alpha,0i\alpha}=0$ for all crystal structures and $\sum_{I} \Phi_{Ii\alpha,0i\beta}=0$ for all centrosymmetric cubic crystal structures. Substituting the latter result into Eq.~\ref{eq:dyn_mat}, we see that there is at least one momentum mode where the diagonal dynamical matrix with respect to ion positions, $D_{i\alpha,i\beta}$, is identically zero. Since the trace of the dynamical matrix is equal to the sum of its eigenvalues, $D_{i\alpha,i\beta}=0$ implies that centrosymmetric cubic crystals are neither stabilized nor destabilized by a harmonic term. For other crystals, we note that the trace of the diagonal dynamical matrix is zero, $\sum_\alpha D_{i\alpha, i\alpha}=0$. Therefore, if some modes are stable ($\omega^2>0$) others will be necessarily unstable ($\omega^2<0$). These results provide strong motivation to revisit the stability of crystals.

In this paper, we study the diagonal matrix of force constants, $\boldsymbol{\Phi}_I\equiv \Phi_{Ii\alpha,0i\beta}$. We focus on the diagonal ($i=j$) elements of the matrix of force constants, since they are sufficient to demonstrate that a system is \emph{not} stable (see Sec.~SII)\footnote{This quantity would not be sufficient to quantify whether a system is stable.}. Moreover, in cases where $\boldsymbol{\Phi}_I=\mathbf{0}$, we additionally examine the symmetry-contracted fourth-order diagonal force constant matrix, $\mathbf{X}_i\equiv X_{Ii\alpha,0i\beta}$ (see Sec.~SIII)\footnote{Higher-order (in)stabilities are, in all cases, weaker than at lower orders.}.

With the strategy and motivation in place, we still face the challenge of calculating the energy. We therefore turn to three limits where we can make progress: the bare ionic crystal, and subsequently the density tight-binding and density nearly-free electron models.


\section{Bare ionic model}
\label{sec:icm}

We start with the simplest system that demonstrates the concept of this paper. For the bare ionic crystal, we consider a one-component crystal of Coulomb point charges of equal sign and infinite extent. This is typical of the systems studied in plasma physics~\cite{Mavadia13}, albeit without a background of positive charges. Working in atomic units, the Coulomb potential is $V(\mathbf{R})=|\mathbf{R}|^{-1}$ corresponding to repulsive interactions between the point charges. Our strategy is to demonstrate that cubic crystals can have a zero matrix of force constants with respect to the motion of a single ion, and therefore centrosymmetric cubic crystals are not necessarily stabilized or destabilized at harmonic order.

In Sec.~\ref{subsec:icmdiscussion}, we discuss the background and key developments in the field of Coulomb crystals, and in Sec.~\ref{subsec:icmanalysis} we analyze our numerical results.

\subsection{Background}
\label{subsec:icmdiscussion}
%

Coulomb crystals are defined by the dominant role of the Coulomb interaction and the simple form of their constituents~\cite{Bonitz08}. In this paper, we consider a special type of `transient Coulomb' crystal, categorized as an unconfined and infinite, one-component system with repulsive interactions. However, the study of Coulomb crystals extends beyond this limiting case and has a history spanning over a century~\cite{Madelung1918}. 

The earliest study of a one-component system was by Madelung in 1918, where he showed that an infinite array of point charges can form an ordered state~\cite{Madelung1918}. Two decades later, Wigner predicted, in his seminal paper, that the electron jellium in metals can form a body-centered cubic crystal at sufficiently low densities~\cite{Wigner34,Wigner38}. The subsequent numerical and experimental confirmation of Wigner crystals sparked interest in the condensed matter community, and a plethora of papers on the general theory~\cite{Trail03,Solanpaa11,Bonsall77,Silvestrov17,Hinarejos12,Coldwell60,Carr61} and stability~\cite{Goldoni96,Goldoni96_2,Palanichamy01,Zucker91} of these systems followed, including detailed quantum Monte Carlo simulations~\cite{Drummond04,Zhu95,Ceperley80,Ceperley78,Candido04,Ortiz99,Ciccariello09,Liu19}. From the plasma physics perspective on the other hand, interest in strongly coupled plasmas, i.e.~plasmas where the average Coulomb energy of a particle is much greater than its average kinetic energy~\cite{Ichimaru82}, led to the prediction that three-dimensional, one-component Coulomb plasmas can also form a body-centered cubic crystal at sufficiently \emph{high} densities and/or low temperatures~\cite{Dubin99}. It was subsequently realized that these two conclusions could be reconciled as opposite density limits of the same problem~\footnote{Additionally, the successful quest to determine the intermediate physics arguably inspired the first application of importance-sampled diffusion Monte Carlo~\cite{Ceperley80}.}. All of these models, however, include a homogeneous positive background of charges to stabilize the system. Indeed, there are two ways to stabilize a repulsive Coulomb crystal: a homogeneous oppositely-charged background, or confinement~\cite{Bonitz08}.

Work on confined plasmas has been performed in a variety of contexts~\cite{Ichimaru82}. Most notably, the structure and Madelung energy~\cite{Hasse91}, as well as the melting of ordered states~\cite{Schiffer02} in spherical Coulomb crystals has been studied in the last thirty years. These systems can also be probed and manipulated experimentally using ions confined to Penning~\cite{Mavadia13} or Paul~\cite{Hornekaer01} traps, with motivation provided by the recent discovery of crystalline plasmas of dust particles in astrophysics~\cite{Chu94}; as well as the industrial success of quantum dot technology~\cite{Ashoori96}. For all of these confined systems, however, the resulting crystal structure is strongly dependent on the shape of the trap~\cite{Schiffer02}. Therefore, no general statements can be made about the equilibrium structure.

In this section, we study the instability of unconfined Coulomb crystals, which we stabilize in later sections through the inclusion of an oppositely-charged background.

\subsection{Analysis}
\label{subsec:icmanalysis}

We first calculate the BO energy of a lattice of ions. We examine the Bravais lattices: simple cubic (cub), body-centred cubic (bcc), and face-centred cubic (fcc). Additionally, we study the diamond (dia) lattice structure, from the fcc family, separately, as it is of special interest due to its extreme material properties, such as hardness and thermal conductivity. We also include the hexagonal close-packed (hcp) and double hexagonal close-packed (dhcp) structures in our analysis, from the hexagonal Bravais lattice family, due their ubiquity in nature (see Sec.~SIV). 

In order to perform the summation over lattice sites in this section we use a rotationally-symmetric summation scheme. We start by defining all unit cells with an atom at the origin and then incrementally add atoms in concentric shells. The long-range contribution is incorporated using the classical Ewald method~\cite{Ewald21} and we compute this summation until convergence to the desired precision. The full details of the numerical model are discussed in Sec.~SV.

\begin{table}
	\begin{ruledtabular}
		\begin{tabular}{c c c c}
			Crystal & $a \boldsymbol{\Phi}^\text{i-i}_{0}$  & $\prescript{}{\text{i-i}}{\hat{\mathbf{m}}_2}$ & $\prescript{}{\text{i-i}}{\hat{\mathbf{m}}^{\intercal}_2} \cdot \left( a\boldsymbol{\Phi}^\text{i-i}_0\right)\cdot\prescript{}{\text{i-i}}{\hat{\mathbf{m}}_2}$  \\ 
			\hline
			$C$ & $\mathbf{0}$ & $-$ & $0$ \\
			$H$ & 
			$-\displaystyle{\frac{k}{2}}\begin{pmatrix}
			1 & 0 & 0 \\
			0 & 1 & 0 \\
			0 & 0 & -2
			\end{pmatrix}$
			& $\pm\hat{\mathbf{e}}_\mathrm{z}$ & $\begin{aligned} k_\text{hcp} &= -0.33 \\ k_\text{dhcp}&= -0.8 \end{aligned}$
		\end{tabular}
	\end{ruledtabular}
\caption{\label{tbl:icm} Diagonal matrices of force constants and minimizing directions for the ion-ion interaction expansion about equilibrium, at second order with lattice spacing, $a$. The cubic crystals are denoted by $C\in\{\text{cub, bcc, fcc, dia}\}$ and the hexagonal crystals by $H\in\{\text{hcp, dhcp}\}$. $\boldsymbol{\Phi}_0$ is the Hessian; $\hat{\mathbf{m}}_2$ is the normalized eigenvector corresponding to the lowest eigenvalue of the Hessian; and $\hat{\mathbf{m}}^{\intercal}_2 \cdot\boldsymbol{\Phi}_0\cdot\hat{\mathbf{m}}_2$ is the projection of the Hessian in the minimizing direction. All values are given to the precision up to which they have converged, or three significant figures, whichever is lower.}
\end{table}

\begin{figure}
\begin{minipage}[b]{\linewidth}
\begin{minipage}[b]{.5\linewidth}
\begin{tikzpicture}
\node at (0,0) {\centering\includegraphicsgood[width=.95\linewidth]{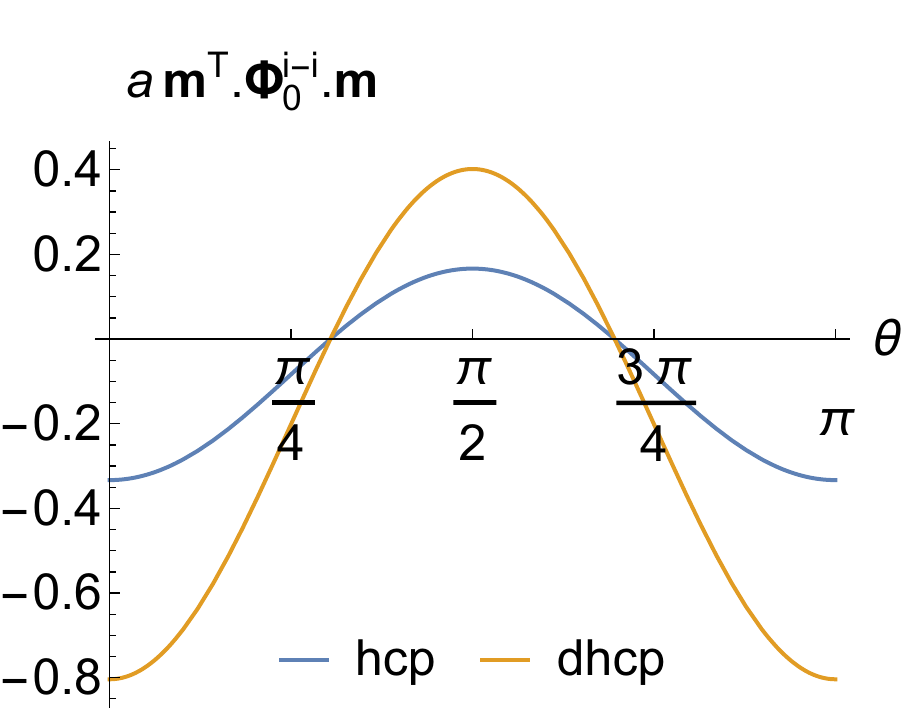}};
\node[overlay] at (-2,1.3) {(a)};
\end{tikzpicture}
\end{minipage}%
\begin{minipage}[b]{.5\linewidth}
\begin{tikzpicture}
\node at (0,0) {\centering\includegraphicsgood[width=.95\linewidth]{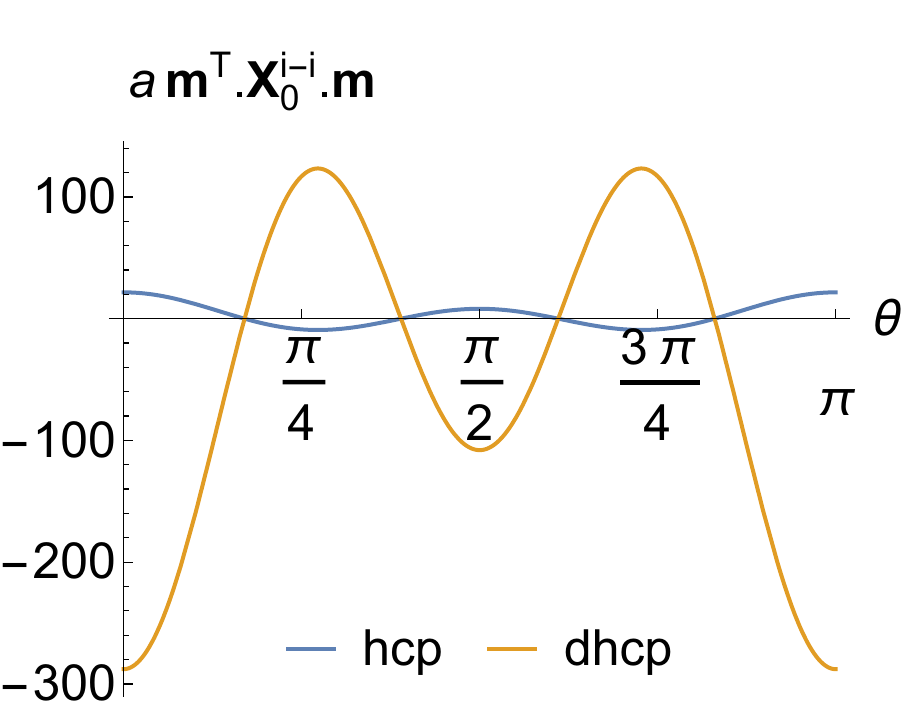}};
\node[overlay] at (-2,1.3) {(b)};
\end{tikzpicture}
\end{minipage}%
\end{minipage}
\caption{Angular variation in the inner product of the matrices of force constants with a direction vector $\mathbf{m}$ at $\phi=\pi/2$ and unit radius, in units of $a^{-1}$. Plots are shown for the (a) second-order, and (b) fourth-order terms for the hcp and dhcp crystal structures. Note that the inner product of the diagonal matrices of force constants is azimuthally symmetric for these systems. The minimizing directions are recorded in Tables~\ref{tbl:icm} \&~\ref{tbl:icm4}.}
\label{fig:hexspheres}
\end{figure}

The diagonal matrices of force constants, directions of greatest instability, and minimal eigenvalues for these crystals are shown in Table~\ref{tbl:icm}. For all crystal structures, the trace of the diagonal matrix of force constants is zero. For cubic systems (cub, bcc, fcc, dia), the diagonal harmonic term is identically zero, whereas for hexagonal systems (hcp, dhcp), the diagonal force constant matrices are indefinite, which implies the system is at a saddle point. We also see that hexagonal structures are stable to diagonal perturbations in the xy-plane, but most unstable to diagonal perturbations in the z-direction, as illustrated in Fig.~\ref{fig:hexspheres}a. The dhcp system is more unstable than the hcp system with this metric due to the higher density of ions.

\begin{table*}
	\begin{ruledtabular}
		\begin{tabular}{c c c c c}
			Crystal & $a\mathbf{X}^\text{i-i}_0$ & $\prescript{}{\text{i-i}}{\hat{\mathbf{m}}_4}$ & ${\prescript{}{\text{i-i}}{\hat{\mathbf{m}}^{\circ 2}_4}}^{\intercal}\cdot\left(a\mathbf{X}^\text{i-i}_0\right)\cdot\prescript{}{\text{i-i}}{\hat{\mathbf{m}}^{\circ 2}_4}$ \\ 
			\hline
			cub &
			$74.6\begin{pmatrix}
			1 & -1.5 & -1.5 \\
			-1.5 & 1 & -1.5 \\
			-1.5 & -1.5 & 1
			\end{pmatrix}$ & 
			$\frac{1}{\sqrt{3}}(\pm\hat{\mathbf{e}}_\mathrm{x}\pm\hat{\mathbf{e}}_\mathrm{y}\pm\hat{\mathbf{e}}_\mathrm{z})$ & $-49.7$ \\
			bcc/fcc/dia &
			$k\begin{pmatrix}
			1 & -1.5 & -1.5 \\
			-1.5 & 1 & -1.5 \\
			-1.5 & -1.5 & 1
			\end{pmatrix}$ & 
			$\{\pm\hat{\mathbf{e}}_\mathrm{x},\pm\hat{\mathbf{e}}_\mathrm{y},\pm\hat{\mathbf{e}}_\mathrm{z}\}$ & $\begin{aligned} k_\text{bcc} &= -74.6 \\ k_\text{fcc}&= -181 \\ k_\text{dia}&= -2570 \end{aligned}$  \\
			hcp & 
			$8.1\begin{pmatrix}
			1 & 1 & 0 \\
			1 & 1 & 0 \\
			0 & 0 & 0
			\end{pmatrix}
			+
			21.6\begin{pmatrix}
			0 & 0 & -1.5 \\
			0 & 0 & -1.5 \\
			-1.5 & -1.5 & 1
			\end{pmatrix}$ &
			$\theta=0.857,\pi - 0.857$ & $-9.3$  \\
			dhcp & 
			$-108\begin{pmatrix}
			1 & 1 & 0 \\
			1 & 1 & 0 \\
			0 & 0 & 0
			\end{pmatrix}
			-
			288\begin{pmatrix}
			0 & 0 & -1.5 \\
			0 & 0 & -1.5 \\
			-1.5 & -1.5 & 1
			\end{pmatrix}$ &
			$\pm\hat{\mathbf{e}}_\mathrm{z}$ & $-288$  \\
		\end{tabular}
	\end{ruledtabular}
	\caption{\label{tbl:icm4} Diagonal force constant matrices and minimizing directions for the ion-ion interaction expansion about equilibrium, at fourth order, with the same conventions as Table~\ref{tbl:icm}. Fourth-order matrices are symmetry contracted as described in Sec.~SIII.}
\end{table*}

\begin{figure}
\begin{minipage}[b]{\linewidth}
\begin{minipage}[b]{.5\linewidth}
\begin{tikzpicture}
\node at (0,0) {\centering\includegraphicsgood[width=.75\linewidth]{cubii4_shrink.pdf}};
\node[overlay] at (-1.9,1.6) {(a)};
\label{fig:cubii2}
\end{tikzpicture}
\end{minipage}%
\begin{minipage}[b]{.5\linewidth}
\begin{tikzpicture}
\node at (0,0) {\centering\includegraphicsgood[width=.75\linewidth]{bccii4_shrink.pdf}};
\node[overlay] at (-1.9,1.6) {(b)};
\label{fig:cubii2}
\end{tikzpicture}
\end{minipage}%
\end{minipage}
\begin{minipage}[b]{\linewidth}
\begin{minipage}[b]{.5\linewidth}
\centering\includegraphicsgood[width=.85\linewidth]{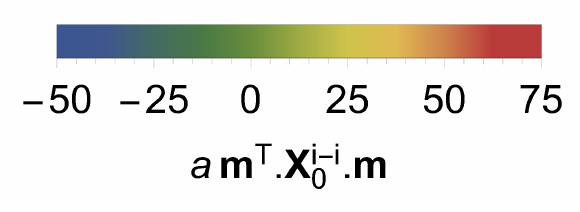}
\end{minipage}%
\begin{minipage}[b]{.5\linewidth}
\centering\includegraphicsgood[width=.85\linewidth]{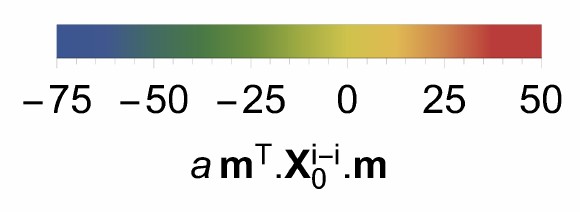}
\end{minipage}%
\end{minipage}
\caption{Angular variation in the inner product of the fourth-order diagonal matrices of force constants with a direction vector $\mathbf{m}$ at unit radius, in units of $a^{-1}$. Plots are shown for the (a) cub, and (b) bcc crystal structures. The plots for the fcc and dia crystal structures have an identical form to (b) with different scales. The scales and minimizing directions are recorded in Table~\ref{tbl:icm4}.}
\label{fig:spheres}
\end{figure}
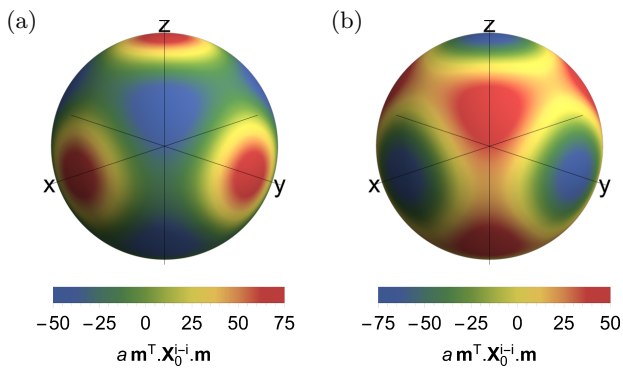

Having found that the cubic crystal stability test can be inconclusive at second order, we turn to a higher-order expansion. An analogous table for the fourth-order diagonal matrices of force constants is shown in Table~\ref{tbl:icm4}~\footnote{Odd power terms in the potential trivially vanish due to symmetry.}. At this order, the cubic systems do not have vanishing contributions, instead they demonstrate a fourth-order instability. Note that the form of the fourth-order matrices is similar in each case, with a varying pre-factor. Plots of the angular variation of these fourth-order matrices are shown in Fig.~\ref{fig:spheres}. As for the hexagonal systems at second order, the system is again at a saddle point. In this case, the configuration is stable to perturbations in the Cartesian basis directions for cub; and in the diagonal directions for bcc, fcc and dia crystals, and visa versa. For completeness, we show that the fourth-order matrices for the hexagonal systems are also indefinite, as shown in Fig.~\ref{fig:hexspheres}b. In the dhcp case, the minimizing directions are again $\pm\hat{\mathbf{e}}_\text{z}$, whereas for the hcp system the most unstable directions have now shifted to $\theta=0.857,\pi-0.857$. The angular variation for the hexagonal systems is rotationally symmetric about the z-axis, since the x- and y-eigenvalues are the same. Note that since higher-order (in)stabilities are always weaker than lower orders, it is unnecessary to examine the higher-order terms for these hexagonal systems. As seen for the second-order case, the magnitude of the instabilities is determined by the ion density. 

This lack of a diagonal harmonic contribution to the energy in cubic systems appears to contradict the 1904 quasi-harmonic model for a crystal of atoms connected by springs~\cite{Drude1904_1, Drude1904_2, Cassidy}. However, as stated before, it is a natural consequence of Gauss' theorem $(\partial_{xx} +\partial_{yy}+\partial_{zz})E^\text{BO}=0$. In a system with cubic symmetry all terms in Gauss' theorem must be identical so each must be zero, $\partial_{\alpha\alpha}E^\text{BO}=0$. Furthermore, $\partial_{\alpha\beta}E^\text{BO}=0$ for these examples by symmetry. Conversely, in systems without cubic symmetry, we can say that if in some direction the second derivative is positive, then in others it must be negative to satisfy Gauss' theorem, and so will never be stable by geometry. The changing sign in the fourth-order derivative in cubic systems is expected as Gauss' theorem requires the net electron flux through a closed surface to be zero, so positive contributions must be counterbalanced by negative contributions. We therefore deduce that it is inevitable that crystalline solids are not stabilized at any order by contributions from the ion-ion potential.  

Note that in this section we have considered a one-component ionic crystal without a neutralizing background to show that cubic structures have the weakest (fourth-order) instability with respect to the motion of a single ion. In Sec.~\ref{sec:nfem} we will show that if a constant neutralizing background is introduced, this would provide a quadratic restoring potential for the ions, which would compensate for this instability. This holds even for non-cubic systems, since it can be shown that the stabilizing contribution to the dynamical matrix from the constant uniform background is greater than the destabilizing contribution from the purely repulsive ionic crystal.


\section{Density tight-binding model}
\label{sec:tbm}

We found that the bare crystal of ions is not stable and so, motivated by the need to stabilize the system, we now consider the simplest model to include electrons to bind the ions: the density tight-binding model. The electrons are tightly bound to each nucleus with a spherical effective charge density parameterized by core and valence orbital radii.

We start by analyzing the model and phase diagram in Sec.~\ref{subsec:tbmanalysis}, and then discuss the interpretation in Sec.~\ref{subsec:tbmdiscussion}.

\subsection{Analysis}
\label{subsec:tbmanalysis}
%

In the density tight-binding approximation, the electrons are situated directly on top of and nearby to the ions. We consider ions that have only spherically symmetric (s-type) orbitals, with the electron density distribution:
\begin{equation*}
\rho_\text{E}(\mathbf{r};c,a_\text{e})\propto \frac{1}{1+\exp\left( \frac{2(|\mathbf{r}|-c)}{a_\text{e}}\right)},
\end{equation*}
where the normalization factor to give net charge neutrality is given in Sec.~SVI A 1. Here $\mathbf{r}$ denotes the displacement of the electron relative to the origin of its associated ion, and $c, a_\text{e}$ characterize the core and valence orbital radii, respectively. The factor of two ensures that the associated wavefunction, defined by $\rho=|\Psi|^2$, reduces to the hydrogenic atom solution, $\sim\exp(-|\mathbf{r}|/a_\text{e})$, in the extreme density tight-binding approximation: $c\ll a_\text{e} \ll a$. We choose this form of the electron orbital density~\cite{Ortiz99}, because it is analytically well behaved for the required derivation and has the correct scaling behavior (see Sec.~SVI A 1). Throughout our calculations, we work to leading order in the density tight-binding approximation. In practice, this implies results up to first order in the small core radial parameter $(c/a_\text{e})$ and second order in the valence radial parameter $(a_\text{e}/a)$.

As mentioned in Sec.~\ref{sec:theory}, to compute the total energy, $E$, we first solve the electronic problem with ions assumed fixed and then we allow the ions to move in the Born-Oppenheimer potential energy surface to account for the ionic contribution to the kinetic energy.  

There are two contributions to the electronic kinetic energy: the energy due to confinement and the energy due to tunneling. We note that the expectation value of the total electronic kinetic energy due to confinement is effectively independent of atom positions, since each potential well in the vicinity of an ion is approximately the same shape. Moreover, the contribution from the electrons tunneling into neighboring wells is exponentially small. Therefore, the expectation value of the total electronic kinetic energy is constant with respect to atom configurations.

We calculate the ion-ion, electron-ion, and electron-electron contributions to the potential energy based on the electron orbital ansatz up to the approximations detailed above. We subsequently add on the contribution to the energy due to the Pauli repulsion of the overlapping electron orbitals, evaluated at the optimal effective radius of atoms in a spherical packing. Finally, we relax the crystal structure to find the optimal lattice constant, $a$. We perform the calculation for each of the crystal structures: cub, bcc, fcc, dia, hcp, and dhcp. 

For both the kinetic and potential energies, we use the same rotationally-symmetric summation scheme for the crystal introduced in Sec.~\ref{sec:icm}. The details of the numerical model are discussed in Sec.~SVI.     

\begin{figure}
		\begin{tikzpicture}
		\node at (0,0) {\centering\includegraphicsgood[width=.95\linewidth]{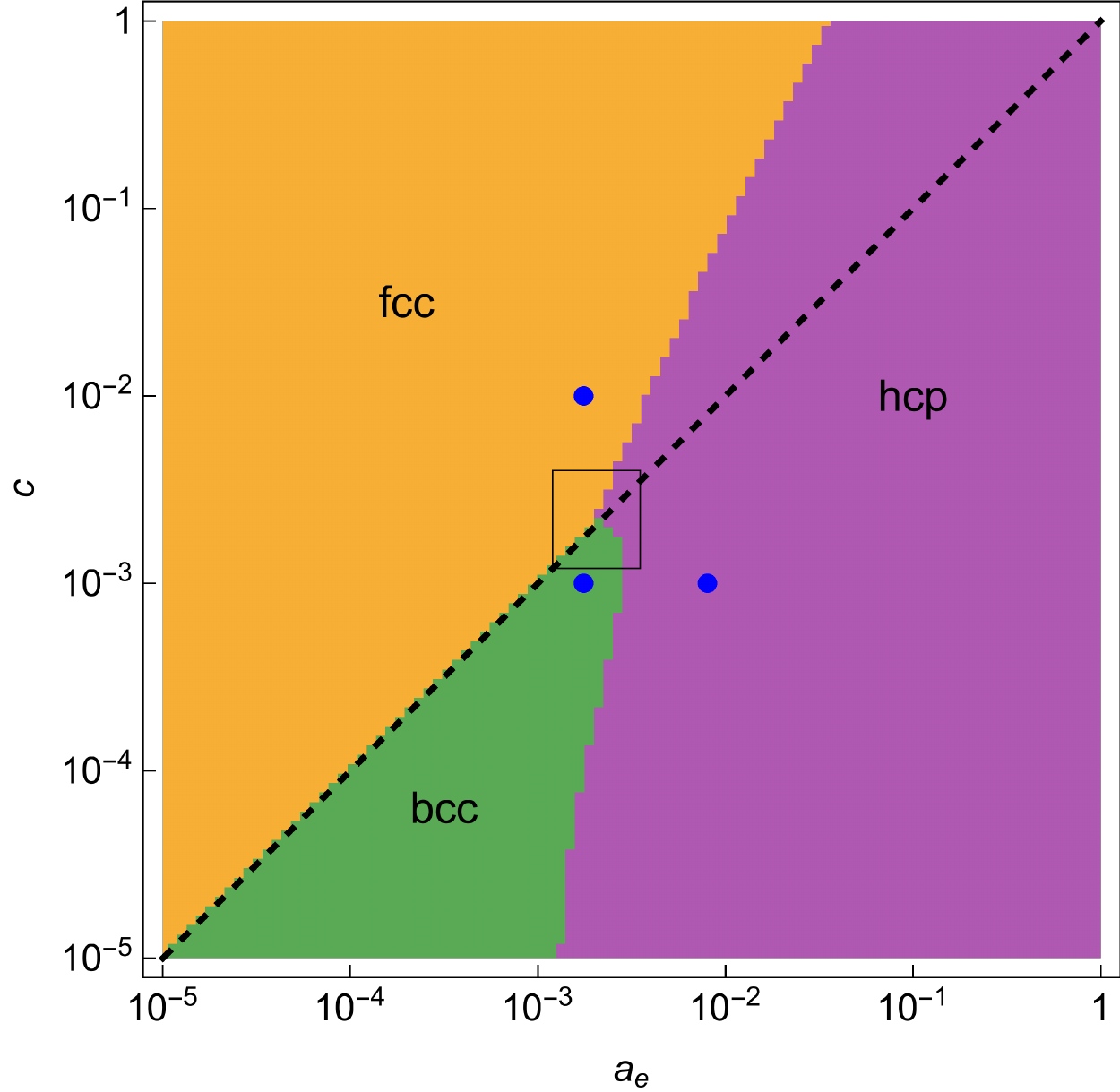}};
		\node[overlay] at (2.5,-1.5) {\centering\includegraphicsgood[width=.25\linewidth]{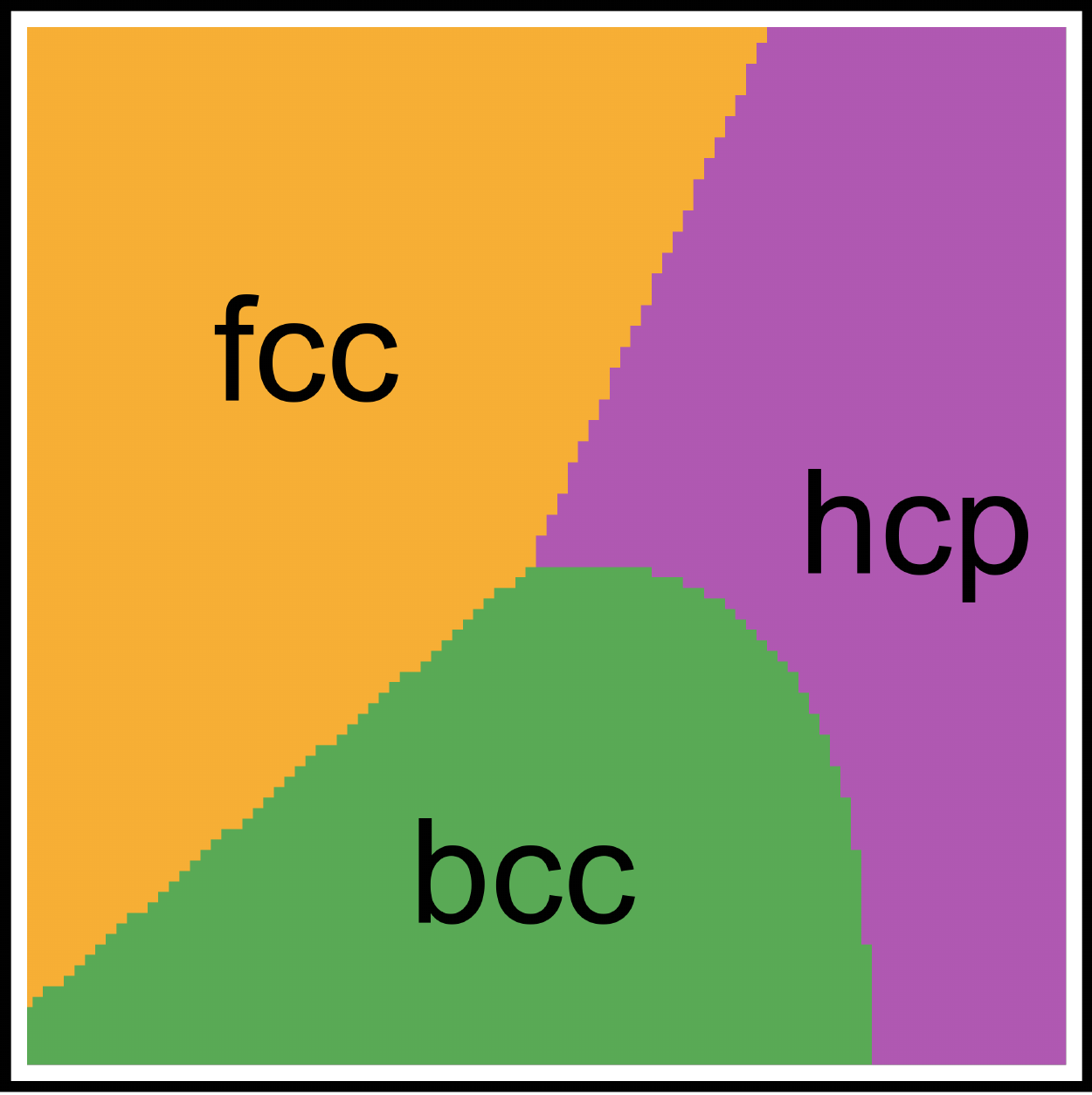}};
		\end{tikzpicture}
		\begin{tikzpicture}[overlay]
		\draw (-4.36,3.95) -- (-2.9,1.53);
		\draw (-3.75,4.66) -- (-0.75,3.7);
		\end{tikzpicture}
	\caption{Phase diagram of the lowest-energy crystal structure out of \{cub, bcc, fcc, dia, hcp, dhcp\} at the optimum lattice constant, summed out to eight shells. The black line separates the valid region for the density tight-binding model: the lower right-hand triangle at $c<a_\text{e}$. The blue points, $\{(1.75\times10^{-3},10^{-2}),(1.75\times 10^{-3},10^{-3}),(8\times 10^{-3},10^{-3})\}$, are analyzed in Fig.~\ref{fig:points}. Inset: Higher-resolution plot of the region enclosed by the black square, highlighting the tricritical point. The diagrams are plotted to a resolution of $100^2$ points.}
	\label{fig:phase_diagram}
\end{figure}
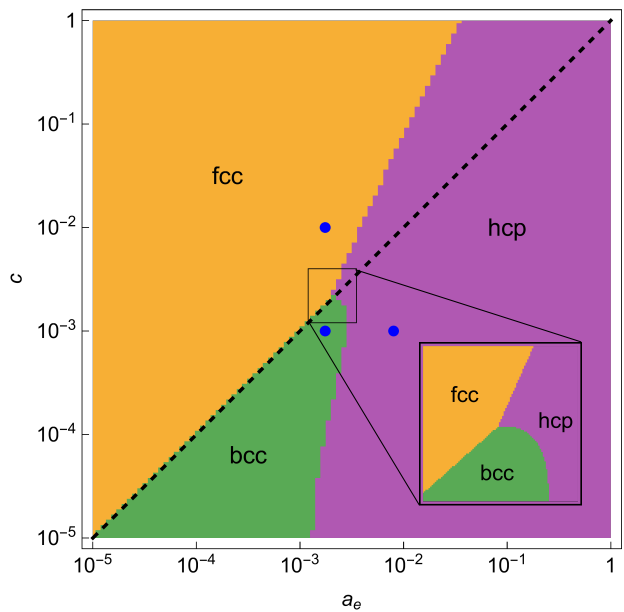

In Fig.~\ref{fig:phase_diagram}, we show the phase diagram of the stable crystal structures with the lowest energy out of the cub, bcc, fcc, dia, hcp, and dhcp lattices. We note that all of the crystal structures are stable with respect to their dynamical matrices in this model and so the preferred crystal structure is determined by the total energy hierarchy. Out of the six crystal structures considered, the hcp, fcc, and bcc structures are found to be the preferred phases. We present a higher-resolution close-up of the tricritical point in the inset of Fig.~\ref{fig:phase_diagram} to analyze the features of interest. The tricritical point is at $(a_\text{e},c)=(2.04,2.13)\times 10^{-3}$ with three transition lines: fcc-bcc at $c\propto a_\text{e}$; fcc-hcp at $c\propto a_\text{e}^{2.5}$; and bcc-hcp at $c=2.13\times10^{-3}$ in the vicinity of the tricritical point. Since all phase transitions between allotropes of crystal structures are first order, the tricritcal point is valid with respect to the vertex rule. Note that other than the restriction imposed by the density tight-binding approximation, in this context $c\lesssim a_\text{e}$, the phase diagram may be extended in both directions.

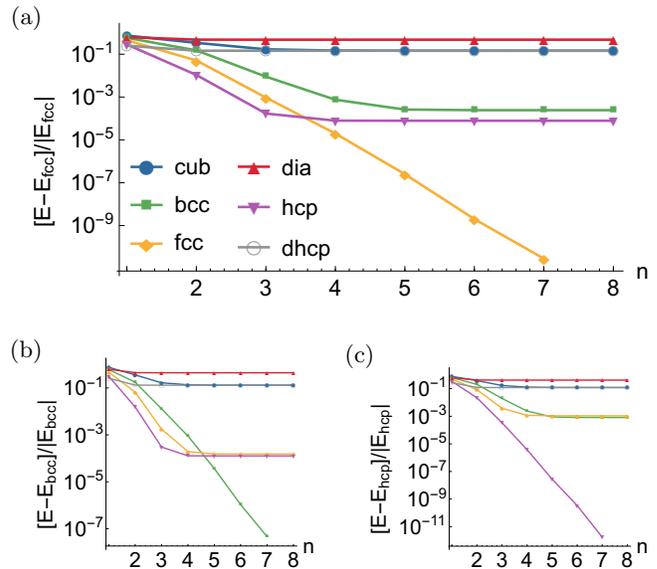
\begin{figure}
\begin{minipage}[b]{\linewidth}
	\begin{tikzpicture}
	\node at (0,0) {\centering\includegraphicsgood[width=.95\linewidth]{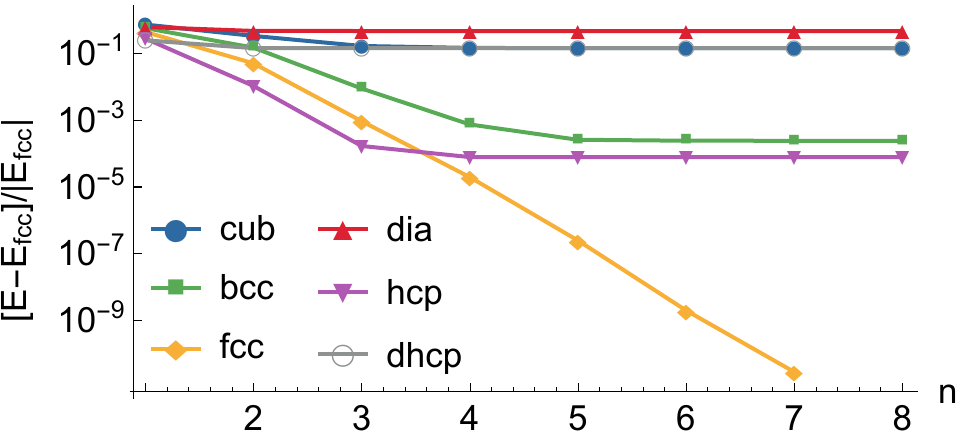}};
	\node[overlay] at (-4.2,1.9) {(a)};
	\end{tikzpicture}
\end{minipage}
\begin{minipage}{\linewidth}
	\vspace{1em}
\end{minipage}
\begin{minipage}[b]{\linewidth}
	\centering
\begin{minipage}[b]{.45\linewidth}
\begin{tikzpicture}
\node at (0,0) {\centering\includegraphicsgood[width=.95\linewidth]{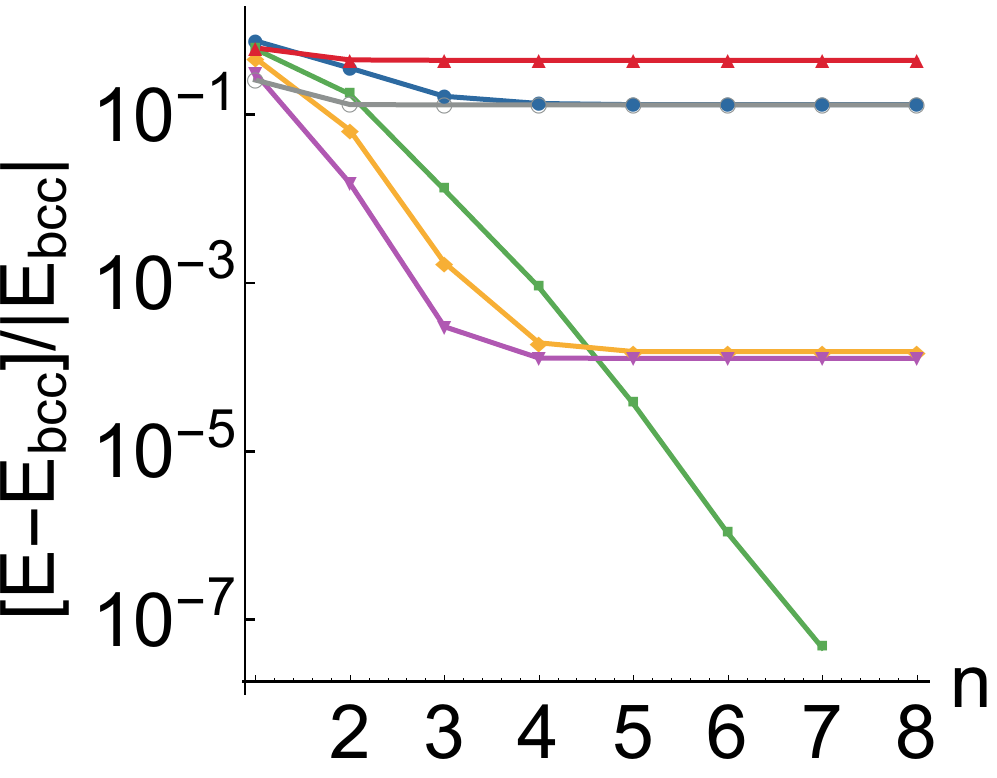}};
\node[overlay] at (-2,1.5) {(b)};
\end{tikzpicture}
\end{minipage}
\hspace{1em}
\begin{minipage}[b]{.45\linewidth}
\begin{tikzpicture}
\node at (0,0) {\centering\includegraphicsgood[width=.95\linewidth]{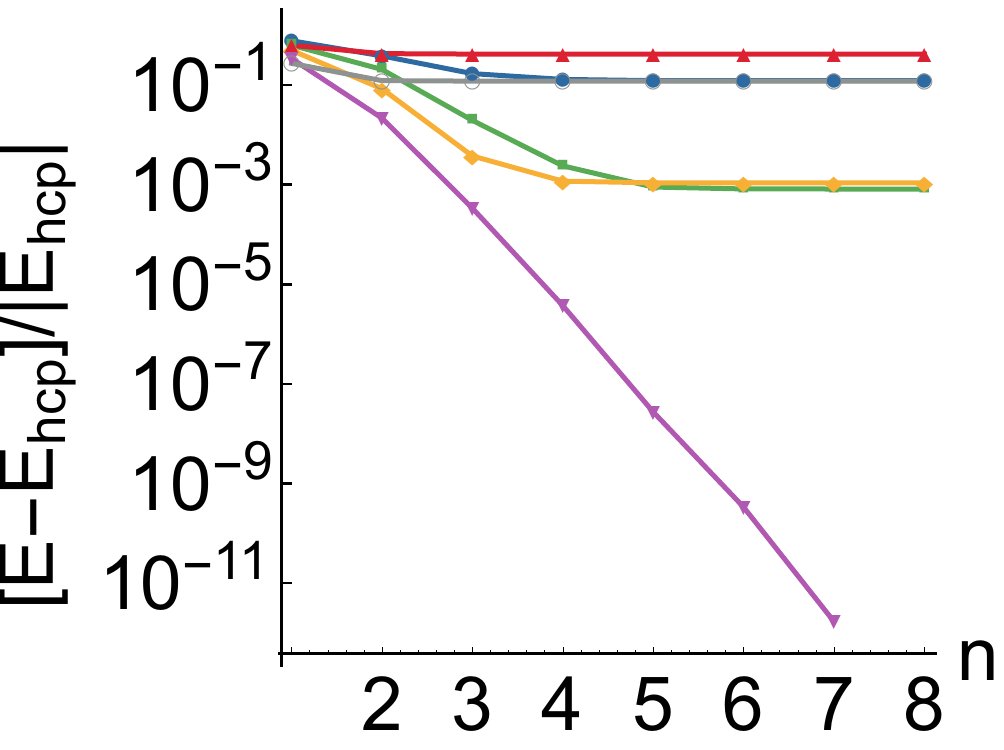}};
\node[overlay] at (-2,1.5) {(c)};
\end{tikzpicture}
\end{minipage}
\end{minipage}
\caption{Detailed analysis of the blue points depicted in Fig.~\ref{fig:phase_diagram}, such that (a) is at $(a_\text{e},c)=(1.75\times10^{-3},10^{-2})$, (b) is at $(1.75\times10^{-3},10^{-3})$, and (c) is at $(8\times10^{-3},10^{-3})$. The plots show the fractional deviation of the energies from the lowest energy value at eight shells, $[E-E_\text{c.s.}(n=8)]/|E_\text{c.s.}(n=8)|$, against the number of shells in the summation, $n$.}
\label{fig:points}
\end{figure}

Now that we have constructed the phase diagram, we verify the convergence of our calculations. Figure~\ref{fig:points} shows a detailed analysis of the blue points depicted in Fig.~\ref{fig:phase_diagram}. Most importantly, we see from plots of the total energy against number of shells of ions in the summation, that convergence is reached at approximately five shells. Therefore, we plot the phase diagram by summing over eight shells, deep into the converged region. We consistently observe that \{bcc, fcc, hcp\} forms the energetically favorable subset of crystal structures.


\subsection{Discussion}
\label{subsec:tbmdiscussion}
%

In this section, we progressed from the ionic crystal in Sec.~\ref{sec:icm} by introducing tightly-bound electrons to stabilize the system. Three phases emerged with noticeably lower energy: hcp, bcc, and fcc. Each is the ground state in different limits, and have been separately noted before, hence the density tight-binding model presented allows us to reconcile previous findings in a unified framework. We now discuss how these phases emerge in the three separate limits of our density tight-binding model. 

It has been known for a long time that three-dimensional Coulomb crystals have a bcc symmetry~\cite{Dubin99}, where the term ``Coulomb crystal" in plasma physics refers to strongly-coupled charged particles with a neutralizing background~\cite{Bonitz08}. In the density tight-binding limit ($c\ll a_\text{e}\ll a$), this is effectively equivalent to the system presented in Fig.~\ref{fig:phase_diagram}. The particle interactions are Coulomb-like, since the effect of the well is still minimal, and the presence of the electrons provides the neutralizing background, albeit highly concentrated around the ions. Therefore, it is unsurprising that we see the same bcc ground state crystal structure. This also has parallels to a Wigner crystal, where the decay of the electronic wavefunction is sufficiently slow to stabilize the crystal~\cite{Wigner34}. 

As soon as we move into the region where $c>a_\text{e}$, we modify the effective interaction through screening. In this region we observe the behavior of screened Coulomb charges, and when $c\gg a_\text{e}$ and $c \gg a$ we observe the density nearly-free electron model. Indeed, it has been shown by Hamaguchi et al.~\cite{Hamaguchi97} that three-dimensional Yukawa crystals have a bcc and fcc phase. They show that there exist two solid phases for the Yukawa crystal: bcc at small screening parameter and a transition to fcc when the screening parameter is increased, which corresponds to moving vertically upwards in our phase diagram.
\begin{figure}
	\centering\includegraphicsgood[width=\linewidth]{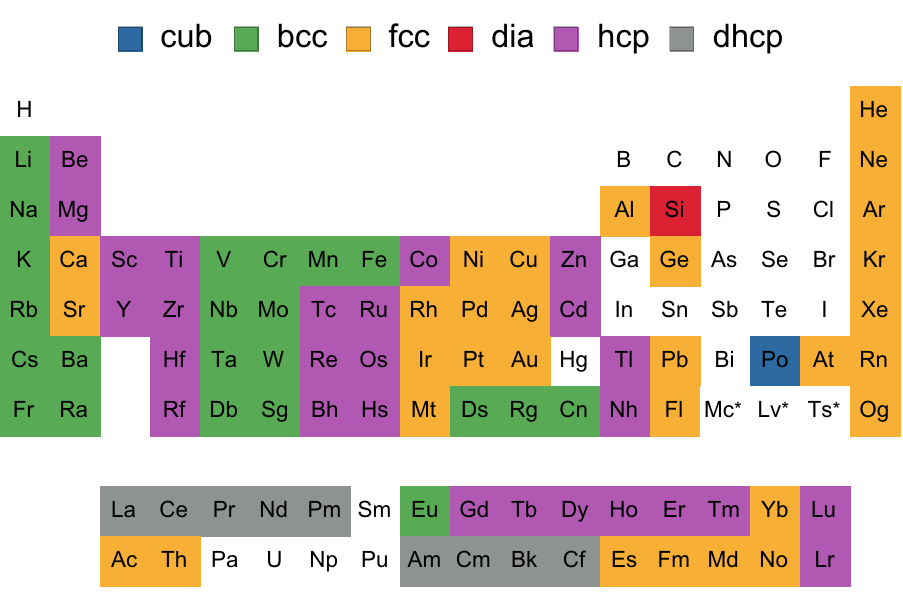}
	\caption{The periodic table of elements labeled according to the crystal structure of their thermodynamically most stable allotrope~\cite{Greenwood97}. The white cells correspond to elements with a crystal structure that is not in the set \{cub, bcc, fcc, dia, hcp, dhcp\}. The crystal structures of elements marked with an asterisk is not known.}
	\label{fig:periodic_table}
\end{figure}

In addition to these extreme limits, our model provides insight into the transition from extreme matter to real materials. From Fig.~\ref{fig:phase_diagram}, we can see that as $a_\text{e}$ is increased, that is the valence electron radius is increased and the density tight-binding approximation is relaxed, the hcp structure is energetically favorable, for both $c<a_\text{e}$ and $c>a_\text{e}$. This shows that in many materials the crystal lattice begins to favor high symmetry and packing factor. The limit applies to many of the Lanthanides and Actinides that are in the tight-binding regime~\cite{Booth01}. Moreover, as shown in Fig.~\ref{fig:periodic_table} and discussed in Sec.~SIV, the hexagonal structure is the most common Bravais lattice in the periodic table. More generally, the vast majority of the periodic table is composed of the bcc, fcc, and hcp crystal structures\footnote{see Sec.~SIV}, identified here are the three most energetically favorable structures.


\section{Density nearly-free electron model}
\label{sec:nfem}

In contrast to the density tight-binding model, where the Bohr radii of the atoms are much smaller than the inter-atomic spacing, we now consider the opposite ``density nearly-free electron" limit, where the Bohr radii mostly overlap. This is applicable to a variety of simple metals in the periodic table, and particularly the Alkali metals. In the weak binding, or density nearly-free electron model, we perform first-order perturbation theory about the jellium model, where the electron density is uniform. Our strategy is to focus on the ion motions found in Sec.~\ref{sec:icm} to be not governed by a harmonic potential, and then investigate whether the electron cloud can compensate for this. The details of the electron cloud densities in this model are presented in Sec.~SVII.

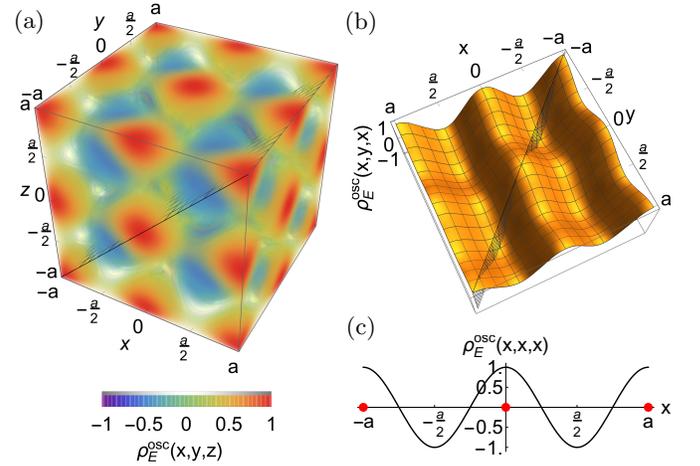
\begin{figure}
\begin{minipage}[b]{\linewidth}
\begin{minipage}[b]{.49\linewidth}
\begin{tikzpicture}
\node at (0,0) {\centering\includegraphicsgood[width=\linewidth]{nfem_3D_shrink.pdf}};
\node[overlay] at (-2,2.8) {(a)};
\label{fig:nfem_3D}
\end{tikzpicture}
\end{minipage}\hfill
\begin{minipage}[b]{.49\linewidth}
\begin{tikzpicture}
\node at (0,0) {\centering\includegraphicsgood[width=\linewidth]{nfem_2D_shrink.pdf}};
\node[overlay] at (0,-3) {\centering\includegraphicsgood[width=\linewidth]{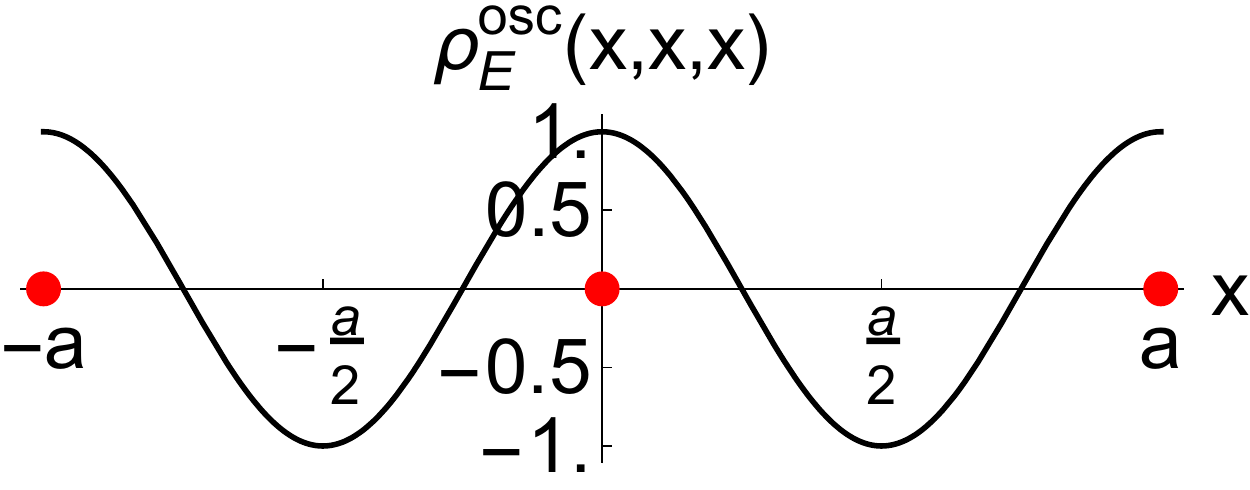}};
\node[overlay] at (-2,1.96) {(b)};
\node[overlay] at (-2,-2.1) {(c)};
\label{fig:nfem_2D}
\end{tikzpicture}
\vspace{5.25em}
\end{minipage}%
\end{minipage}
\caption{(a) Three-, (b) two-, and (c) one-dimensional plots of the oscillatory part of the electron cloud density, $\rho^\text{osc}_\text{E}(x,y,z)=[\cos(k x) +\cos(k y) + \cos(k z)]/3$ with $k=2\pi/a$, for the simple cubic lattice in the density nearly-free electron model. (a) Color and opacity both denote the magnitude of $\rho^\text{osc}_\text{E}(x,y,z)$. (b) Plot of the meshed cross-section depicted in (a), through the density extrema. (c) Plot of the meshed cross-section depicted in (b), through the density extrema. The red points illustrate the positions of the ions.}
\label{fig:nfem}
\end{figure}

The density nearly-free electron model comprises a lattice of ions with Coulomb repulsion, as studied in Sec.~\ref{sec:icm}, together with an oscillatory and near-uniform electron cloud density, $\rho_\text{E}$. The expectation value of the total electronic kinetic energy in this model is therefore directly proportional to the Fermi energy. In accordance with first-order perturbation theory, this electron cloud density may be split into two parts, $\rho_\text{E}=\rho^\text{cst}_\text{E}+\rho^\text{osc}_\text{E}$, with $\rho^\text{cst}_\text{E}$ corresponding to the constant jellium-like density, and $\rho^\text{osc}_\text{E}$ corresponding to the oscillatory density due to the electron-ion interaction and the geometry of the ionic lattice. An example of the oscillatory electron cloud density for the simple cubic lattice is shown in Fig.~\ref{fig:nfem}. In each case, we ensure that the density range is normalized such that $\max(\rho^\text{osc}_\text{E})=u$, where $u$ is the oscillation strength, and that the integral of $\rho^\text{osc}_\text{E}$ over a unit cell is equal to zero.

In order to calculate the total diagonal force constant matrix, we proceed by summing the ion-ion, electron-ion, and electron-electron contributions from the BO potential. For the ion-ion contribution, we take results directly from Sec.~\ref{sec:icm}. Note that when performing the real-space summation over shells for the ion-ion contribution, the zeroth-order contribution to the BO energy is divergent, whereas the matrix of force constants converges. In fact, there are divergent zeroth-order contributions for the electron-ion and electron-electron contributions too, corresponding to the jellium-like term in the electron density. These divergent terms cancel, which is reflected in the Ewald summation. However, we gain additional insight by directly calculating the diagonal force constants matrix in each case. The constant electron-ion contribution to the diagonal force constants matrix must satisfy $\sum_\alpha \Phi_{0i\alpha, 0i\alpha}^\text{e-i,cst} = 4\pi\rho^\text{cst}_\text{E}$ by Poisson's law. The uniform neutralizing background stabilizes any crystal with respect to diagonal force constants and its contribution is summarized in Table~\ref{tbl:nfem}.

The oscillatory electron-ion contribution to the BO energy, $E^\text{BO,osc}_\text{e-i}(\mathbf{u})=-2\sum_{I} \int V_\text{i}(\mathbf{R}_I-\mathbf{u}+\mathbf{r}_\text{e}) \rho^\text{osc}_\text{E}(\mathbf{r}_\text{e})\mathrm{d}\mathbf{r}_\text{e}$, may be simplified by noting that all ions are equivalent and so we can focus on the ion at the origin. Subsequently calculating the energy per atom allows us to drop the summation over ions and write
\begin{equation}
\label{eq:nfemei}
E^\text{BO,osc}_\text{e-i}(\mathbf{u})=-2 \int V_\text{i}(-\mathbf{u}+\mathbf{r}_\text{e}) \rho^\text{osc}_\text{E}(\mathbf{r}_\text{e})\,\mathrm{d}\mathbf{r}_\text{e},
\end{equation}
where $V_\text{i}$ is given by the Coulomb potential and the oscillatory part of the electron density is approximated by a cosine function (see Sec.~SVII).

Finally, for the electron-electron contribution, $E_\text{e-e}^\text{BO}=\sum_{I} \iint V_\text{e}(\mathbf{R}_I-\mathbf{u}+\mathbf{r}_\text{e}-\mathbf{u}_\text{e}) \rho_\text{E}(\mathbf{r}_\text{e}) \rho_\text{E}(\mathbf{u}_\text{e}) \mathrm{d}\mathbf{r}_\text{e} \mathrm{d}\mathbf{u}_\text{e}$, we may drop the summation by the same argument. Note also that the electron potential does not depend on the displacement of the central ion. Hence, excluding the zeroth-order term and working to first-order in the perturbation strength, we may write the oscillatory contribution to the electron-electron BO energy as
\begin{equation}
\label{eq:nfemee}
E^\text{BO,osc}_\text{e-e}=2 \rho^\text{const}_\text{E} \int_{\mathbf{r}_\text{e}\in\substack{\text{unit}\\\text{cell}}} \int_{\mathbf{u}_\text{e}\in\mathbb{R}^3}  V_\text{e}(\mathbf{r}_\text{e}-\mathbf{u}_\text{e})  \rho^\text{osc}_\text{E}(\mathbf{u}_\text{e})\,\mathrm{d}\mathbf{u}_\text{e} \mathrm{d}\mathbf{r}_\text{e},
\end{equation}  
where the factor of two is from the addition of both cross terms, and $V_\text{e}$ is again given by the Coulomb potential. For all structures the positive and negative regions of the oscillating electron density $\rho_\text{E}^\text{osc}$ cancel and therefore the oscillatory electron-electron contribution is zero, $E^\text{BO,osc}_\text{e-e}=0$.

The summation of the leading-order terms from Sec.~\ref{sec:icm}, the jellium contribution, as well as the oscillatory contributions from Eqs.~\ref{eq:nfemei} and~\ref{eq:nfemee} yields the total BO energy, the diagonal matrix of force constants, and hence an instability discriminant.

\begin{table}
	\begin{ruledtabular}
		\begin{tabular}{c c}
			Crystal & $a(\boldsymbol{\Phi}^\text{e-i,cst}_0 + \boldsymbol{\Phi}^\text{e-i,osc}_0)$ \\ 
			\hline
			$C$ & $\displaystyle{\frac{4\pi}{3}\left(k + \sqrt{2\pi}\tilde{u}\right)\mathbf{I}}$ \\
			hcp & 
			$\displaystyle{\frac{4\sqrt{2}\pi}{41}\left(1 + \sqrt{\pi}\tilde{u}\right)}\begin{pmatrix}
			16 & 0 & 0 \\
			0 & 16 & 0 \\
			0 & 0 & 9
			\end{pmatrix}$ \\
			dhcp
			& $\displaystyle{\frac{4\sqrt{2}\pi}{697}\left(2 + \sqrt{\pi}\tilde{u}\right)}\begin{pmatrix}
			218 & 0 & 0 \\
			0 & 218 & 0 \\
			0 & 0 & 261
			\end{pmatrix}$ \\
		\end{tabular}
	\end{ruledtabular}
	\caption{\label{tbl:nfem}Electron-ion contributions to the diagonal force constants matrix in the density nearly-free electron model. The cubic crystals are denoted by $C\in\{\text{cub, bcc, fcc}\}$. The prefactors for the constant electron-ion contributions for cubic systems are $\{k_\text{cub},k_\text{bcc},k_\text{fcc}\}=\{1,2,4\}$. $\mathbf{I}$ and $\tilde{u}$ denote the identity matrix and dimensionless oscillation strength, respectively.}
\end{table}

The harmonic energy contributions for the crystal structures is summarized in Table~\ref{tbl:nfem}. We note that all of the electron-ion contributions are positive at this order. We can see that the cubic structures all have isotropic matrices, whereas the hexagonal structures are only isotropic in the xy-plane, as expected by symmetry.


For all of the crystals considered, the electron-ion term from the constant electron background alone is sufficient to counterbalance the corresponding ion-ion term. The oscillating electron background provides additional stability for these diagonal terms. We note that, for the complete stability hierarchy in the density nearly-free electron system, the dynamical matrices need to be studied, as well as additional effects, such as the electron per atom concentration and the band lowering at the Brillouin zone boundaries~\cite{Kittel86, Esther08}. 

In the empty lattice approximation ($\tilde{u}\approx0$), we find that cubic systems have positive diagonal harmonic terms of larger magnitude than hexagonal systems, and in particular the fcc structure has the largest diagonal harmonic term. This potentially concurs with the nearly-free limit $c \gg a_\text{e}$ and $c \gg a$ of the density tight-binding model. Furthermore, this matches observations in the periodic table, not only for the quintessential empty lattice case study: aluminium~\cite{Walter60}, but also nickel, copper, silver, and gold~\cite{Choy00} -- all of which have an fcc structure.     

In this section, we have shown that, considering diagonal harmonic terms with respect to the motion of a single ion, all ionic crystals are counterbalanced with the addition of a constant neutralizing background, and that a first-order oscillatory component to the background does not have a destabilizing effect. The fcc lattice has the largest such term, in agreement with many itinerant elemental solids. 


\section{Conclusion}
\label{sec:conclusion}

In this paper, we have studied lattice instability of unconfined crystal structures at zero-temperature in the bare ionic, density tight-binding, and density nearly-free electron models. We analyzed the diagonal matrices of force constants to expose instability and focused on the \{cub, bcc, fcc, dia, hcp, dhcp\} structures due to their prevalence in nature and distinctive properties. 

In Sec.~\ref{sec:icm}, we studied a bare one-component system of point Coulomb charges. First, we reviewed the history of the field, and noted that the bcc structure is special for being the stable crystal structure for both the low density Wigner crystal and the high density Coulomb crystal in the one-component plasma model. We then demonstrated that in this regime, centrosymmetric cubic crystal structures can have no diagonal contribution to the dynamical matrix at quadratic order but instead an instability at fourth order, whereas all other crystal structures have an instability at second order. This is counter to the paradigmatic quasi-harmonic model of ions connected by springs~\cite{Drude1904_1, Drude1904_2, Cassidy}. These findings motivated us to continue and examine the preferred structure as we permit background charges to stabilize the system. 

In Sec.~\ref{sec:tbm}, we stabilized the lattice through the addition of electron orbitals. We constructed a density tight-binding model, and found that in the extreme density tight-binding limit, the bcc structure is the most stable, as suspected from the results and discussion in Sec.~\ref{sec:icm}. We also showed that if we tune the parameters to increase screening in our pseudopotential model of the nucleus and approach the nearly-free electron limit, the fcc structure is the stable ground state. This agrees with theoretical studies of unconfined three-dimensional Yukawa crystals in the literature. Finally, we report the second dominant phase to be hcp as we tune away from density tight-binding, which accords to trends in the periodic table. The use of the density tight-binding model with the systematic analysis of terms has allowed us to combine the emergence of these three separate phases into a single framework. 

In Sec.~\ref{sec:nfem}, we briefly examined the instability of crystal structures in the opposite limit, density nearly-free electrons, which is representative of many common metals. In this model, we found that the instability of every crystal structure is counterbalanced with the addition of a constant neutralizing background, and that a first-order oscillatory perturbation to the background does not have a destabilizing effect. We note here that a formal stability analysis would require the complete dynamical matrix. The most stable crystal structure in the empty lattice approximation according to this metric is fcc, agreeing not only with a limit of the density tight-binding model, but also with several structures observed in the periodic table.  

By investigating three simple cases, motivated by the absence of diagonal force constants in cubic Coulomb crystals, this paper highlights the connections and limitations of paradigmatic crystal models for stability. For many real materials, there are a plethora of important effects that need to be taken into consideration to determine the optimal lattice structure e.g.~temperature effects, the shape of the atomic orbitals, the precise band structure, or the van der Waals interaction, which leaves scope for future work. However, we have identified minimal models to illustrate the interesting physical effects at play, shown how the density tight-binding and density nearly-free electron theories may be linked, and connected historical theories of crystal structures at low energies. We hope that this closer look at the energies and force constants for these three models will instill a greater appreciation and understanding of the requirements for crystal stability, as well as its connection with lattice geometry.


\begin{acknowledgments}
We thank Emilio Artacho, Edward Linscott, Chris Pickard, Daniel Rowlands, Be\~{n}at Mencia Uranga, Claudio Castelnovo, Yu~Yang~Fredrik~Liu, and Pablo~L\'opez~R\'ios for useful discussions. B.A. acknowledges support from the Engineering and Physical Sciences Research Council under Grant No. EP/M506485/1 and the Swiss National Science Foundation under Grant No. PP00P2\_176877. G.J.C. acknowledges support from the Royal Society. Open access to the data in this paper is available at the DSpace@Cambridge repository.
\end{acknowledgments}

\bibliographystyle{apsrev4-1}
\bibliography{ms}

\end{document}



\onecolumngrid


\begin{center}
	\textbf{\large Supplementary Material: Absence of diagonal force constants in cubic Coulomb crystals}
\end{center}

\tableofcontents

\setcounter{equation}{0}
\setcounter{figure}{0}
\setcounter{table}{0}
\setcounter{section}{0}
\makeatletter
\renewcommand{\theequation}{S\arabic{equation}}
\renewcommand{\thefigure}{S\arabic{figure}}
\renewcommand{\thetable}{S\arabic{table}}
\renewcommand{\thesection}{S\Roman{section}}
\renewcommand{\bibnumfmt}[1]{[S#1]}
\renewcommand{\citenumfont}[1]{S#1}

\section{Proof that $\sum_\alpha \Phi_{Ii\alpha,0i\alpha}=0$ for $I\neq 0$}
\label{sec:proof}

The diagonal matrices of force constants, $\boldsymbol{\Phi}_I\equiv\Phi_{Ii\alpha,0i\beta}$, corresponding to the diagonal dynamical matrix, $D_{i\alpha,i\beta}$, are computed by Taylor expanding a crystal with respect to atom perturbations, as discussed in Sec.~I. In order to demonstrate Poisson's law, $\sum_\alpha \Phi_{0i\alpha,0i\alpha}=0$, it is sufficient to perturb only the central atom, since $\sum_\alpha \Phi_{0i\alpha,0i\alpha}$ corresponds to the Laplacian. In order to show $\sum_\alpha \Phi_{Ii\alpha,0i\alpha}=0$ for $I\neq0$, however, we need to subsequently perturb a second atom to compute the mixed second derivatives $\Phi_{Ii\alpha,0i\alpha}$.

For symmetry, let us consider a crystal of atoms consisting of superimposed positive and negative charges, and choose to perturb the positive charges. After displacing the central positive charge by $\mathbf{u}_1=(x_1,y_1,z_1)$, we subsequently displace a positive charge at $\mathbf{R}_I=(X,Y,Z)$ by $\mathbf{u}_2=(x_2,y_2,z_2)$. The negative shadow charges remain fixed. In this notation, $\mathbf{R}_I$ denotes the separation vector between unit cells $0$ and $I$ in the crystal. Examining the symmetric interaction between these two atoms (four charges), yields the potential energy contribution
\begin{equation*}
E_I=\frac{1}{|\mathbf{R}_I+(\mathbf{u}_2-\mathbf{u}_1)|}-\frac{1}{|\mathbf{R}_I+\mathbf{u}_2|}-\frac{1}{|\mathbf{R}_I-\mathbf{u}_1|}+\frac{1}{|\mathbf{R}_I|}=\frac{1}{2}\mathbf{u}_1^\intercal\boldsymbol{\Phi}_I\mathbf{u}_2 + O(\text{cubic}),
\end{equation*}
where
\begin{equation*}
\boldsymbol{\Phi}_{I} =
\frac{2}{(X^2+Y^2+Z^2)^{5/2}}
\begin{pmatrix}
-2X^2 + Y^2 + Z^2 & -3XY & -3XZ \\
-3XY & X^2 - 2Y^2 + Z^2 & -3YZ \\
-3XZ & -3YZ & X^2 + Y^2 - 2Z^2
\end{pmatrix}.
\end{equation*}
The trace of this matrix is zero for any separation vector $\mathbf{R}_I$. Hence, the formula $\sum_\alpha \Phi_{Ii\alpha,0i\alpha}=0$ holds for all integer $I\neq 0$ and any crystal structure.

\section{Higher-order Derivative Test}
\label{sec:hodt}

For a single-variable, real-valued and sufficiently differentiable function, $f$, let the first $(n-1)$ derivatives vanish such that 
\begin{equation*}
f'(c)=\dots=f^{(n-1)}(c)=0 \;\;\; \text{and} \;\;\; f^{(n)}(c)\neq 0,
\end{equation*}
where $c$ is a constant in the domain of the function, and $n\in\mathbb{Z}^{+}$. In this case, the $n$th derivative may be used as a discriminant to determine the nature of the turning points.

If $n$ is even:
\begin{itemize}
\item $f^{(n)}(c)<0$ $\Rightarrow$ $c$ is a local maximum,
\item $f^{(n)}(c)>0$ $\Rightarrow$ $c$ is a local minimum.
\end{itemize} 

If $n$ is odd:
\begin{itemize}
\item $f^{(n)}(c)<0$ $\Rightarrow$ $c$ is a strictly decreasing point of inflection,
\item $f^{(n)}(c)>0$ $\Rightarrow$ $c$ is a strictly increasing point of inflection.
\end{itemize} 
Hence, this test can classify the critical points of $f$ in all cases, provided $f^{(n)}(c)\neq 0$ for some value of $n$~\cite{Gkioulekas14}. 

The higher-order derivative test may be generalized to multi-dimensional problems. Denoting $D^{(p)}f$ as the $p$th-order multivariate derivative of $f$, it can be shown that under corresponding conditions:
\begin{itemize}
\item $D^{(p)}f(c)$ is negative definite $\Rightarrow$ $c$ is a strict local maximum.
\item $D^{(p)}f(c)$ is positive definite $\Rightarrow$ $c$ is a strict local minimum,
\item $D^{(p)}f(c)$ is indefinite $\Rightarrow$ $c$ is a saddle point,
\item $D^{(p)}f(c)$ is zero or semidefinite $\Rightarrow$ the test is inconclusive.
\end{itemize}
Note that, unlike the single-variable test, this test is not conclusive in all cases~\cite{Nerenberg91}.

\section{Higher-order Matrices of Force Constants }
\label{sec:ae}

In this paper we consider the effect of displacing atoms originally at $\{\mathbf{R}_{Ii}^0\}$, on their nearest neighbors, with a energy $E$ and general displacements $\{\mathbf{R}_{Ii}\}$. We may expand the energy such that:
\begin{align*}
E(\{\mathbf{R}_{Ii}\})=&E(\{\mathbf{R}_{Ii}^0\})+\sum_{Ii\alpha} \left.\frac{\partial E}{\partial u_{Ii\alpha}}\right|_{\mathbf{u}=\mathbf{0}}u_{Ii\alpha} +\frac{1}{2!}\sum_{Ii\alpha}\sum_{Jj\beta} \Phi_{Ii\alpha, Jj\beta} u_{Ii\alpha}u_{Jj\beta} \\
&+\frac{1}{3!}\sum_{Ii\alpha}\sum_{Jj\beta}\sum_{Kk\gamma} \left.\frac{\partial^3 E}{\partial u_{Ii\alpha}\partial u_{Jj\beta}\partial u_{Kk\gamma}}\right|_{\mathbf{u}=\mathbf{0}} u_{Ii\alpha} u_{Jj\beta} u_{Kk\gamma}\\
&+\frac{1}{4!}\sum_{Ii\alpha}\sum_{Jj\beta}\sum_{Kk\gamma}\sum_{Ll\delta} X_{Ii\alpha, Jj\beta, Kk\gamma, Ll\delta} u_{Ii\alpha} u_{Jj\beta} u_{Kk\gamma} u_{Ll\delta} + \dots,
\end{align*}
where $\{I,J,K,L\}$ are unit cell indices, $\{i, j, k, l\}$ are basis atom indices, and $\{\alpha, \beta, \gamma, \delta\}$ are Cartesian directions. As stated in the main text, translational invariance allows us to consider $\Phi_{Ii\alpha, 0j\beta}$ and $X_{Ii\alpha, 0j\beta, Kk\gamma, Ll\delta}$ without loss of generality. Furthermore, exploiting the symmetry of the system, we additionally contract the fourth-order matrix of force constants such that $X_{Ii\alpha, 0j\beta, Kk\gamma, Ll\delta}\delta_K^0 \delta_L^K\delta^\alpha_\beta \delta^\gamma_\delta = X_{Ii\alpha, 0j\alpha, 0k\gamma, 0l\gamma}$, which allows us to write the diagonal force constant matrices analogously as $\boldsymbol{\Phi}_I\equiv \Phi_{Ii\alpha, 0i\beta}$ and $\mathbf{X}_I\equiv X_{Ii\alpha, 0i\beta}$. Both of these matrices are symmetric in $(\alpha, \beta)$.

\section{Crystal Structures in the Periodic Table}
\label{sec:csipt}

\begin{figure}
\includegraphicsgood[width=\linewidth]{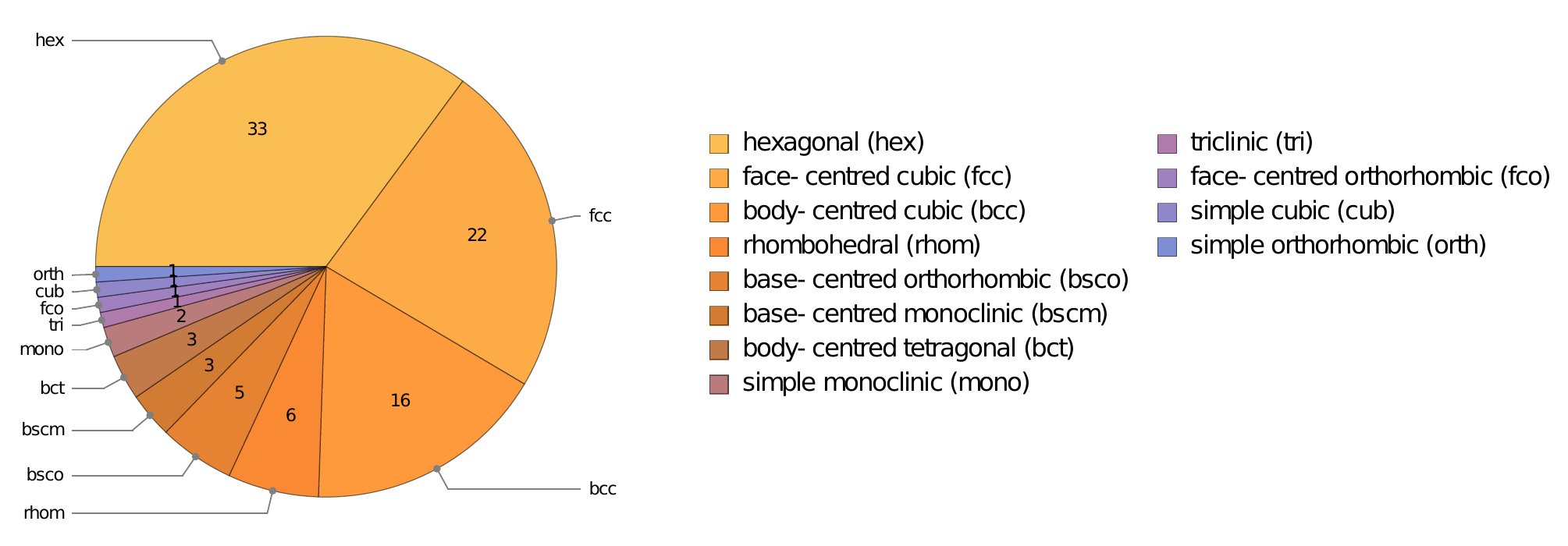}
\caption{\label{fig:pie}Breakdown of crystal structures (by Bravais lattice) for the most thermodynamically stable allotropes of elements in the periodic table, at standard temperature and pressure. The number of elements exhibiting each crystal structure is given in the corresponding section of the chart. Crystal structure data is provided for an unambiguous subset containing 94 out of the 118 elements, using Mathematica's ElementData function~\cite{Mathematica}.}
\end{figure}

Sufficiently stable elements in the periodic table may be grouped in accordance with their crystal structure.  A breakdown of the crystal structures (by Bravais lattice) in the periodic table is presented in Fig.~\ref{fig:pie}. In the cases where an element exhibits multiple crystal structures at standard temperature and pressure, the most thermodynamically stable allotrope is given. 

In three-dimensions, all crystal structures are derived from fourteen possible Bravais lattices. However, some of the derived crystal structures are worth studying separately, either due to their ubiquity (e.g.~in the case of the hcp structure: the most common crystal structure in nature) or interesting properties (e.g.~in the case of diamond). The cub, bcc, fcc, dia, hcp, and dhcp crystal structures are studied in particular in this paper because they only have one free parameter: the lattice constant. Furthermore, this group of crystal structures accounts for approximately three quarters of the known crystal structures in the periodic table.

\section{Numerical Model}
\label{sec:nm}

In this section, we outline the numerical details of how the crystal structure summations were performed. In the interests of clarity, we use simplified notation in this section and consider a one-component crystal with the displacement of a single ion at the origin by a displacement vector $\mathbf{R}$ and total potential energy $E$.  

\subsection{Ewald summation}
\label{subsec:ewald}

Ewald summation is the standard method to compute Coulomb interactions in infinite periodic systems, such as crystals~\cite{Ewald21}. The method works by splitting the Coulomb potential into a singular short-range term, which is evaluated in real space, and a continuous long-range term, which is evaluated in momentum space. The split is performed such that $V(r)=1/r=f(\alpha r)/r -(1-f(\alpha r))/r$, where $f(r)= \erf(r)$ is the error function, $1-f(r)= \erfc(r)$ is the complementary error function, and $\alpha>0$ is the Ewald splitting parameter. The error function is typically chosen because it corresponds to a Gaussian spreading function for the point charges and its Fourier coefficients are known analytically.

In the 3D periodic Coulomb problem, the total potential may be written as
\begin{equation*}
E=\sum_I \sum_{i < j} \frac{1}{|\mathbf{R}_I+\mathbf{r}_{ij}|},
\end{equation*}
where $\mathbf{r}_{ij}\equiv \mathbf{r}_i - \mathbf{r}_j$ and we sum over all distinct pairs of identical unit charges. Rewriting the total potential in the Ewald formalism yields
\begin{equation*}
E=\frac{1}{2}\sum_I {\vphantom{\sum}}' \sum_{i,j} \frac{\erfc(\alpha|\mathbf{R}_I+\mathbf{r}_{ij}|)}{|\mathbf{R}_I+\mathbf{r}_{ij}|} + \frac{1}{2}\sum_{I \neq 0} \sum_{i,j} \frac{4\pi e^{-k_I^2/(4\alpha^2)}}{k_I^2 a^3} e^{-\mathrm{i}\mathbf{k}\cdot \mathbf{r}_{ij}} - \frac{\alpha N_\text{u}}{\sqrt{\pi}}, 
\end{equation*}
where $\mathbf{k}_I=2\pi\mathbf{R}_I/a$ is the momentum, $a$ is the unit cell length, and $N_\text{u}$ is the number of ions per unit cell\footnote{The $\sum '$ denotes that the $i=j$ term is omitted for $I=0$.}. The first term is short-range and evaluated directly by summing concentric shells, as discussed in the following sections. The second term is summed analogously in reciprocal space with a cutoff empirically set such that the Ewald splitting parameter $\alpha\approx2$~\cite{Gibbon02}. The final term is the self energy, which cancels the corresponding contribution from the momentum sum. Depending  on the exact parameters chosen for the cutoffs, the scaling of the classical Ewald summation is between $O(N^{3/2})$ and $O(N^2)$, where $N$ is the number of ions in the system.

\begin{figure}
	\includegraphics[width=\textwidth]{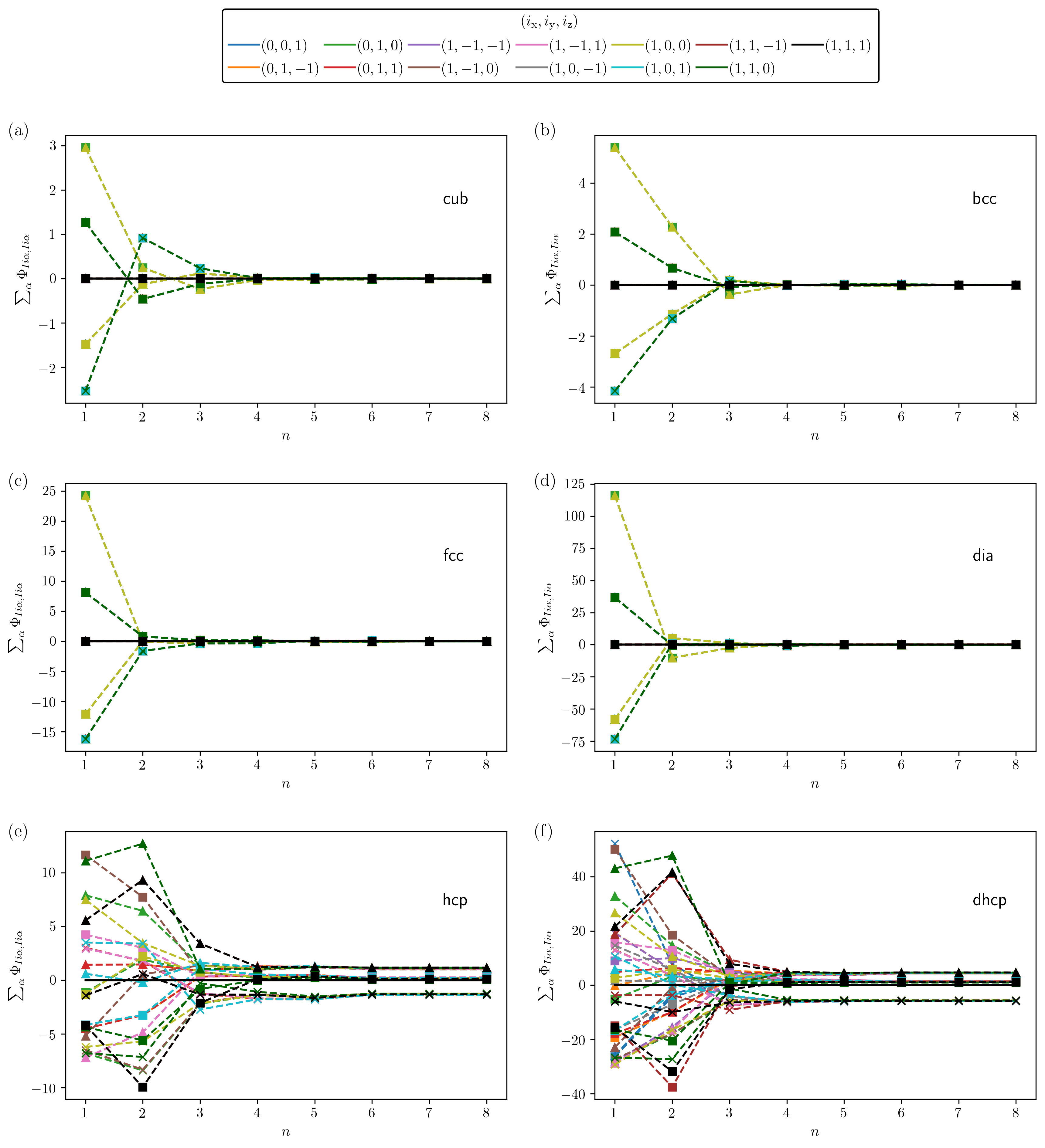}
	\caption{\label{fig:trace}The trace of the diagonal matrix of force constants, $\Phi_{Ii\alpha,Ii\beta}$, for all $\mathbf{R}_I$ satisfying $0<|\mathbf{R}_I|\leq\sqrt{3}$ -- excluding those related by inversion symmetry $\mathbf{R}_I\leftrightarrow-\mathbf{R}_I$ -- against the number of shells in the summation, $n$, for the (a) cub, (b) bcc, (c) fcc, (d) dia, (e) hcp, and (f) dhcp crystal structures. To explicitly show the convergence, we also present the diagonal matrix elements (dashed lines): $\Phi_{Ii0,Ii0}$ (triangles), $\Phi_{Ii1,Ii1}$ (squares), and $\Phi_{Ii2,Ii2}$ (crosses) for each $\mathbf{R}_I$ and crystal structure.}
\end{figure}

To demonstrate the convergence of the classical Ewald method, we show in Fig.~\ref{fig:trace} the computation of the trace of the diagonal matrices of force constants with respect to the motion of a single ion,  $\Phi_{Ii\alpha,Ii\beta}$, against the number of shells in the real-space summation, $n$, for a variety of perturbation centers, $\mathbf{R}_I=i_\mathrm{x}\mathbf{a}+i_\mathrm{y}\mathbf{b}+i_\mathrm{z}\mathbf{c}$, where $\{\mathbf{a},\mathbf{b},\mathbf{c}\}$ are the crystal basis vectors. For cubic crystals, it can be seen that $\sum_\alpha \Phi_{Ii\alpha,Ii\alpha}=0$ by Poisson's law and $\Phi_{Ii\alpha,Ii\alpha}=0$ by symmetry. For hexagonal crystals, we again note that $\sum_\alpha \Phi_{Ii\alpha,Ii\alpha}=0$ by Poisson's law, however the diagonal elements $\Phi_{Ii\alpha,Ii\alpha}\neq0$. In this paper, we sum shells up to and including $n=8$, which is deep into the converged region.

\subsection{Cubic systems (cub, bcc, fcc, dia)}
\label{subsec:cs}

\begin{figure}
\begin{minipage}[b]{\linewidth}
\begin{minipage}[b]{.33\linewidth}
\begin{tikzpicture}
\node at (0,0) {\centering\includegraphicsgood[width=.9\linewidth]{bcc_shrink.pdf}};
\node[overlay] at (-2.8,2.7) {(a)};
\label{fig:cubii2}
\end{tikzpicture}
\end{minipage}%
\begin{minipage}[b]{.33\linewidth}
\begin{tikzpicture}
\node at (0,0) {\centering\includegraphicsgood[width=.9\linewidth]{fcc_shrink.pdf}};
\node[overlay] at (-2.8,2.7) {(b)};
\label{fig:cubii2}
\end{tikzpicture}
\end{minipage}%
\begin{minipage}[b]{.33\linewidth}
\begin{tikzpicture}
\node at (0,0) {\centering\includegraphicsgood[width=.9\linewidth]{dia_shrink.pdf}};
\node[overlay] at (-2.8,2.7) {(c)};
\label{fig:cubii2}
\end{tikzpicture}
\end{minipage}%
\end{minipage}
\caption{\label{fig:structures}Origin-centric unit cells for the (a) bcc, (b) fcc, and (c) dia, crystal structures. These structures have two, four, and eight ions per unit cell, respectively. All lengths are given in units of the lattice constant, and the coloring distinguishes the position along the z-axis. The displacement vectors for these plots are given in Table~\ref{tbl:cubicvecs}a.}
\end{figure}
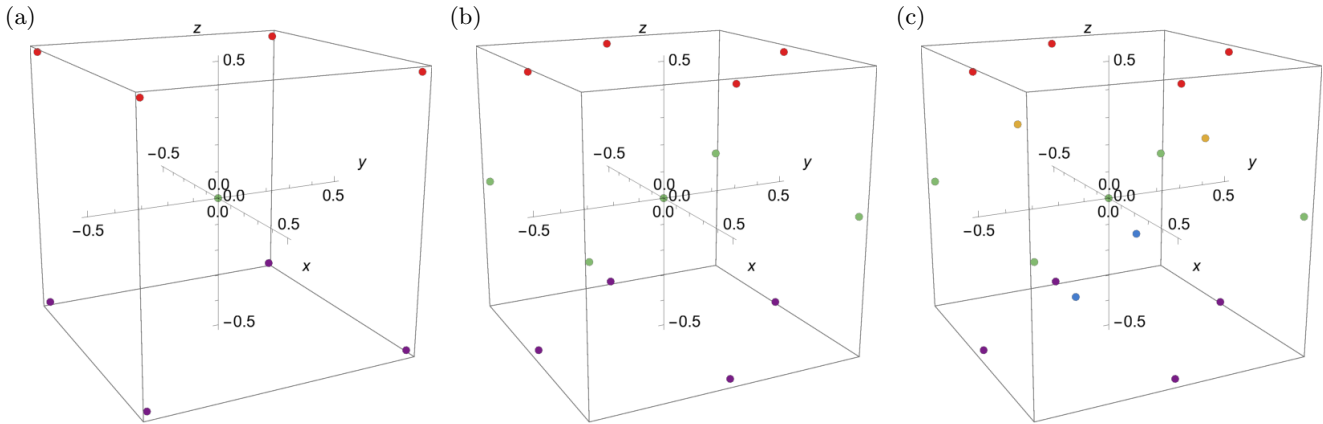

In this paper, we consider unit cells with an ion situated at the origin in all cases. We refer to these as \emph{origin-centric} unit cells, and we choose these in order to minimize finite system size error when summing radially outwards over many shells, as well as to simplify the computations. The unit cell for the simple cubic crystal consists of one ion situated at the origin. The unit cells for the bcc, fcc, and dia crystal lattices are shown in Fig.~\ref{fig:structures}.

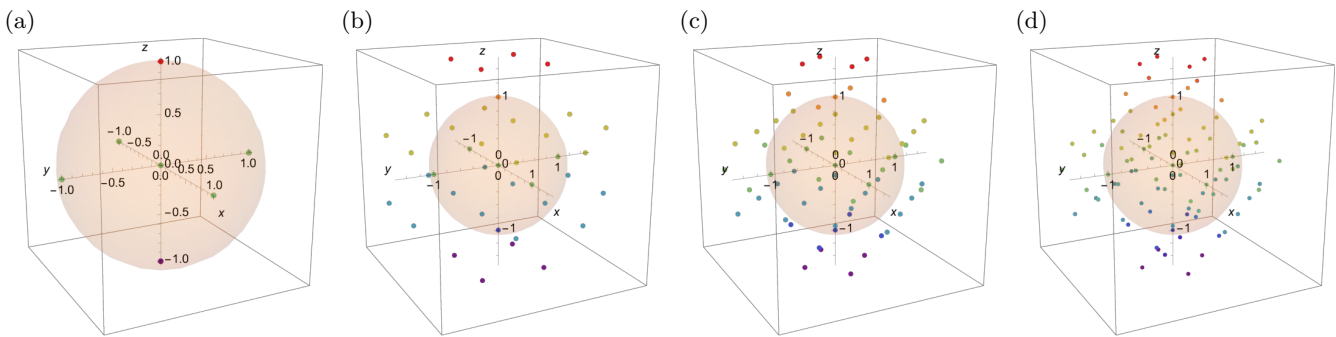
\begin{figure}
\begin{minipage}[b]{\linewidth}
\begin{minipage}[b]{.25\linewidth}
\begin{tikzpicture}
\node at (0,0) {\centering\includegraphicsgood[width=.9\linewidth]{cub_sphere_shrink.pdf}};
\node[overlay] at (-2,2.2) {(a)};
\label{fig:cubii2}
\end{tikzpicture}
\end{minipage}%
\begin{minipage}[b]{.25\linewidth}
\begin{tikzpicture}
\node at (0,0) {\centering\includegraphicsgood[width=.9\linewidth]{bcc_sphere_shrink.pdf}};
\node[overlay] at (-2,2.2) {(b)};
\label{fig:cubii2}
\end{tikzpicture}
\end{minipage}%
\begin{minipage}[b]{.25\linewidth}
\begin{tikzpicture}
\node at (0,0) {\centering\includegraphicsgood[width=.9\linewidth]{fcc_sphere_shrink.pdf}};
\node[overlay] at (-2,2.2) {(c)};
\label{fig:cubii2}
\end{tikzpicture}
\end{minipage}%
\begin{minipage}[b]{.25\linewidth}
\begin{tikzpicture}
\node at (0,0) {\centering\includegraphicsgood[width=.9\linewidth]{dia_sphere_shrink.pdf}};
\node[overlay] at (-2,2.2) {(d)};
\label{fig:cubii2}
\end{tikzpicture}
\end{minipage}%
\end{minipage}
\caption{\label{fig:oneshell} Illustration of the points included in a one-shell summation of the (a) cub, (b) bcc, (c) fcc, and (d) dia, crystal structures. All points from the nearest-neighbor unit cells are plotted. The centers of neighboring unit cells lie within a unit sphere (light orange). All lengths are given in units of the lattice constant, and the coloring of points distinguishes their position along the z-axis.}
\end{figure}

In order to sum to $n$ shells, we include all ions in units cells whose origins are situated within a radius of $n$ lattice constants, as illustrated in Fig.~\ref{fig:oneshell}. We continue to sum in this fashion until the properties of interest, such as the diagonal force constant matrices, converge to the desired precision.

The coordinates of the unit cell sites for these cubic systems is shown in Table~\ref{tbl:cubicvecs}a and the corresponding potentials are given in Table~\ref{tbl:cubicvecs}b. Hence, the summation over $n$ shells may be written explicitly as
\begin{equation*}
E_C=\sum_I V_{C}(\mathbf{R}_I -\mathbf{R})=\sum_{i_\mathrm{x}^2+i_\mathrm{y}^2+i_\mathrm{z}^2\leq n^2} V_C \left(
a\begin{pmatrix}
i_\mathrm{x} \\
i_\mathrm{y} \\
i_\mathrm{z}
\end{pmatrix} 
-
\begin{pmatrix}
X \\
Y \\
Z
\end{pmatrix}
\right)-V\left(
\begin{pmatrix}
X \\
Y \\
Z
\end{pmatrix}
\right),
\end{equation*}
where $C\in\{\text{cub, bcc, fcc, dia}\}$ denotes the cubic crystal structure under consideration, and $\{i_\mathrm{x},i_\mathrm{y},i_\mathrm{z}\}$ are integers. The complete summation, including the long-range contribution, yields the potential energy of displacing the ion at the origin to a position $\mathbf{R}$. The converged expression can then be expanded to quadratic order in $\mathbf{R}$, for example, to extract the diagonal force constant matrix.

\begin{table}
(a) Displacement Vectors
\vspace{0.5em}
\begin{ruledtabular}
\begin{tabular}{c c c}
Crystal & Plane & Displacement Vectors of Sites in the Unit Cell \\
\hline
\multirow{2}{*}{bcc} & $z=a/2$ & $\mathbf{r}^{\text{bcc}}_{1}=\frac{a}{2}(1,1,1)$, $\mathbf{r}^{\text{bcc}}_{2}=\frac{a}{2}(-1,1,1)$, $\mathbf{r}^{\text{bcc}}_{3}=\frac{a}{2}(1,-1,1)$, $\mathbf{r}^{\text{bcc}}_{4}=\frac{a}{2}(-1,-1,1)$, \\
 & $z=-a/2$ & $\mathbf{r}^{\text{bcc}}_{5}=\frac{a}{2}(1,1,-1)$, $\mathbf{r}^{\text{bcc}}_{6}=\frac{a}{2}(1,-1,-1)$, $\mathbf{r}^{\text{bcc}}_{7}=\frac{a}{2}(-1,1,-1)$, $\mathbf{r}^{\text{bcc}}_{8}=\frac{a}{2}(-1,-1,-1)$ \\
\hline
\multirow{3}{*}{fcc} & $z=0$ & $\mathbf{r}^{\text{fcc}}_{1}=\frac{a}{2}(1,1,0)$, $\mathbf{r}^{\text{fcc}}_{2}=\frac{a}{2}(-1,1,0)$, $\mathbf{r}^{\text{fcc}}_{3}=\frac{a}{2}(1,-1,0)$, $\mathbf{r}^{\text{fcc}}_{4}=\frac{a}{2}(-1,-1,0)$, \\
 & $y=0$ & $\mathbf{r}^{\text{fcc}}_{5}=\frac{a}{2}(1,0,1)$, $\mathbf{r}^{\text{fcc}}_{6}=\frac{a}{2}(-1,0,1)$, $\mathbf{r}^{\text{fcc}}_{7}=\frac{a}{2}(1,0,-1)$, $\mathbf{r}^{\text{fcc}}_{8}=\frac{a}{2}(-1,0,-1)$, \\
 & $x=0$ & $\mathbf{r}^{\text{fcc}}_{9}=\frac{a}{2}(0,1,1)$, $\mathbf{r}^{\text{fcc}}_{10}=\frac{a}{2}(0,-1,1)$, $\mathbf{r}^{\text{fcc}}_{11}=\frac{a}{2}(0,1,-1)$, $\mathbf{r}^{\text{fcc}}_{12}=\frac{a}{2}(0,-1,-1)$ \\
\hline
\multirow{6}{*}{dia} & $z=0$ & $\mathbf{r}^{\text{dia}}_{1}=\frac{a}{8}(4,4,0)$, $\mathbf{r}^{\text{dia}}_{2}=\frac{a}{8}(-4,4,0)$, $\mathbf{r}^{\text{dia}}_{3}=\frac{a}{8}(4,-4,0)$, $\mathbf{r}^{\text{dia}}_{4}=\frac{a}{8}(-4,-4,0)$, \\
 & $y=0$ & $\mathbf{r}^{\text{dia}}_{5}=\frac{a}{8}(4,0,4)$, $\mathbf{r}^{\text{dia}}_{6}=\frac{a}{8}(-4,0,4)$, $\mathbf{r}^{\text{dia}}_{7}=\frac{a}{8}(4,0,-4)$, $\mathbf{r}^{\text{dia}}_{8}=\frac{a}{8}(-4,0,-4)$, \\
 & $x=0$ & $\mathbf{r}^{\text{dia}}_{9}=\frac{a}{8}(0,4,4)$, $\mathbf{r}^{\text{dia}}_{10}=\frac{a}{8}(0,-4,4)$, $\mathbf{r}^{\text{dia}}_{11}=\frac{a}{8}(0,4,-4)$, $\mathbf{r}^{\text{dia}}_{12}=\frac{a}{8}(0,-4,-4)$, \\
 & $z=a/4$ & $\mathbf{r}^{\text{dia}}_{13}=\frac{a}{8}(2,2,2)$, $\mathbf{r}^{\text{dia}}_{14}=\frac{a}{8}(-2,-2,2)$, \\
 & $y=a/4$ & $\mathbf{r}^{\text{dia}}_{15}=\frac{a}{8}(-2,2,-2)$, \\
 & $x=a/4$ & $\mathbf{r}^{\text{dia}}_{16}=\frac{a}{8}(2,-2,-2)$
\end{tabular}
\end{ruledtabular}
\vspace{0.5em}
(b) Potentials
\vspace{0.5em}
\begin{ruledtabular}
\begin{tabular}{c c c}
Crystal & Atoms per Unit Cell & Potential \\
\hline
bcc & 2 & $\displaystyle{V_{\text{bcc}}(\mathbf{R})=V(\mathbf{R})+\frac{1}{8}\sum_{i=1}^8  V(\mathbf{R}+\mathbf{r}_i^{\text{bcc}})}$  \\
fcc & 4 & $\displaystyle{V_{\text{fcc}}(\mathbf{R})=V(\mathbf{R})+\frac{1}{4}\sum_{i=1}^{12}  V(\mathbf{R}+\mathbf{r}_i^{\text{fcc}})}$ \\
dia & 8 & $\displaystyle{V_{\text{dia}}(\mathbf{R})=V(\mathbf{R})+\frac{1}{4}\sum_{i=1}^{12}  V(\mathbf{R}+\mathbf{r}_i^{\text{dia}}) + \sum_{i=13}^{16}  V(\mathbf{R}+\mathbf{r}_i^{\text{dia}})}$
\end{tabular}
\end{ruledtabular}
\caption{\label{tbl:cubicvecs}(a) Displacement vectors for sites in a unit cell, and (b) corresponding unit cell potentials, for the bcc, fcc, and dia crystal structures. For the displacement vectors, the site at the origin is omitted and all vectors are given in terms of the lattice constant, $a$.}
\end{table}

\subsection{Hexagonal systems (hcp, dhcp)}
\label{subsec:hs}

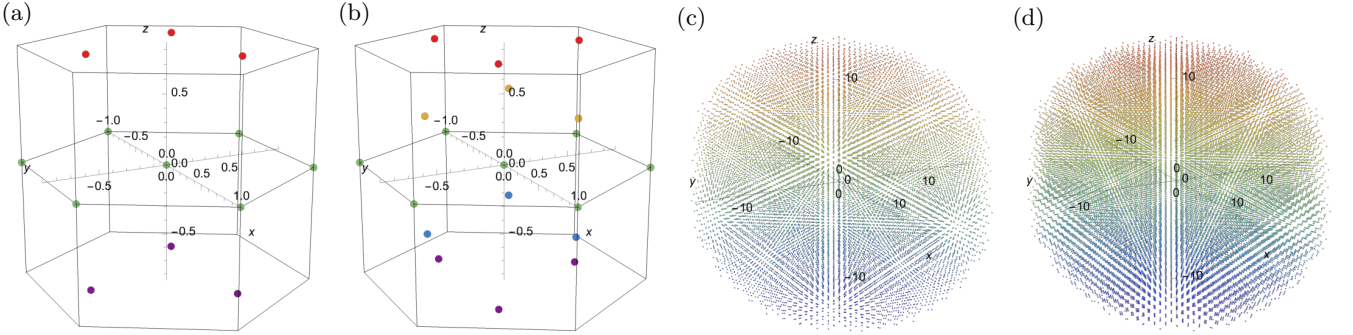
\begin{figure}
\begin{minipage}[b]{\linewidth}
\begin{minipage}[b]{.25\linewidth}
\begin{tikzpicture}
\node at (0,0) {\centering\includegraphicsgood[width=.9\linewidth]{hcp_shrink.pdf}};
\node[overlay] at (-2,2.2) {(a)};
\label{fig:cubii2}
\end{tikzpicture}
\end{minipage}%
\begin{minipage}[b]{.25\linewidth}
\begin{tikzpicture}
\node at (0,0) {\centering\includegraphicsgood[width=.9\linewidth]{dhcp_shrink.pdf}};
\node[overlay] at (-2,2.2) {(b)};
\label{fig:cubii2}
\end{tikzpicture}
\end{minipage}%
\begin{minipage}[b]{.25\linewidth}
\begin{tikzpicture}
\node at (0,0) {\centering\includegraphicsgood[width=.9\linewidth]{hcp_sphere_shrink.pdf}};
\node[overlay] at (-2,2.2) {(c)};
\label{fig:cubii2}
\end{tikzpicture}
\end{minipage}%
\begin{minipage}[b]{.25\linewidth}
\begin{tikzpicture}
\node at (0,0) {\centering\includegraphicsgood[width=.9\linewidth]{dhcp_sphere_shrink.pdf}};
\node[overlay] at (-2,2.2) {(d)};
\label{fig:cubii2}
\end{tikzpicture}
\end{minipage}%
\end{minipage}
\caption{\label{fig:hexstruc}[(a), (b)] Origin-centric unit cells for the (a) hcp, and (b) dhcp, crystal structures. These structures have six and twelve ions per unit cell, respectively. [(c), (d)] Illustrations of the (c) hcp, and (d) dhcp, crystal structures plotted up to eight shells. All lengths are given in units of the lattice constant, and the coloring distinguishes the position along the z-axis. The displacement vectors for these plots are given in Table~\ref{tbl:hexvecs}.}
\end{figure}

Hexagonal systems are treated in an analogous fashion to cubic systems, except now more care is needed since the vectors to neighboring unit cells are not orthogonal. The origin-centric unit cells for the hcp and dhcp crystal lattices are shown in Figs.~\ref{fig:hexstruc}a \&~\ref{fig:hexstruc}b and the corresponding displacement vectors and potentials are presented in Table~\ref{tbl:hexvecs}. Hence for these systems, the (unnormalized) basis set, to go from one unit cell to another, may be denoted as
\begin{equation}
\label{eq:hexbasis}
\{\mathbf{a},\mathbf{b},\mathbf{c}\}=\frac{a}{2}\left\{
\begin{pmatrix}
3 \\
-\sqrt{3} \\
0
\end{pmatrix},
\begin{pmatrix}
3 \\
\sqrt{3} \\
0
\end{pmatrix},
\begin{pmatrix}
0 \\
0 \\
\frac{4\sqrt{6}}{3}
\end{pmatrix}\right\},
\end{equation}
where $a$ is the lattice constant in the xy-plane. In this case, the summation over $n$ shells may be explicitly written as
\begin{equation*}
E_H=\sum_I V_{H}(\mathbf{R}_I -\mathbf{R})=\sum_{\left(\frac{\sqrt{3}}{2}(i_\mathrm{x}+i_\mathrm{y})\right)^2+\left(\frac{i_\mathrm{y}-i_\mathrm{x}}{2}\right)^2+i_\mathrm{z}^2\leq n^2} V_H \left(
\frac{a}{2}\begin{pmatrix}
3(i_\mathrm{x}+i_\mathrm{y}) \\
\sqrt{3}(i_\mathrm{y}-i_\mathrm{x}) \\
\frac{4\sqrt{6}}{3}i_\mathrm{z}
\end{pmatrix} 
-
\begin{pmatrix}
X \\
Y \\
Z
\end{pmatrix}
\right)-V\left(
\begin{pmatrix}
X \\
Y \\
Z
\end{pmatrix}
\right),
\end{equation*}
where $H\in\{\text{hcp, dhcp}\}$ denotes the hexagonal crystal structure under consideration, and $\{i_\mathrm{x},i_\mathrm{y},i_\mathrm{z}\}$ are integers. Figures~\ref{fig:hexstruc}c \&~\ref{fig:hexstruc}d show the sites included in these summations up to eight shells, which is typically the number at which the desired precision converged. Note the approximate spherical symmetry of these systems.

\begin{table}
(a) Displacement Vectors
\vspace{0.5em}
\begin{ruledtabular}
\begin{tabular}{c c c}
Crystal & Plane & Displacement Vectors of Sites in the Unit Cell \\
\hline
    \multirow{4}{*}{hcp}& \multirow{2}{*}{$z=0$}  & $\mathbf{r}^{\text{hcp}}_{1}=a(1,0,0)$, $\mathbf{r}^{\text{hcp}}_{2}=\frac{a}{2}(1,\sqrt{3},0)$, $\mathbf{r}^{\text{hcp}}_{3}=\frac{a}{2}(-1,\sqrt{3},0)$,\\
    & & $\mathbf{r}^{\text{hcp}}_{4}=a(-1,0,0)$, $\mathbf{r}^{\text{hcp}}_{5}=\frac{a}{2}(-1,-\sqrt{3},0)$, $\mathbf{r}^{\text{hcp}}_{6}=\frac{a}{2}(1,-\sqrt{3},0)$, \\
    & $z=2\sqrt{6}/3$ & $\mathbf{r}^{\text{hcp}}_{7}=\frac{a}{6}(3,\sqrt{3},2\sqrt{6})$, $\mathbf{r}^{\text{hcp}}_{8}=\frac{a}{6}(-3,\sqrt{3},2\sqrt{6})$, $\mathbf{r}^{\text{hcp}}_{9}=\frac{a}{3}(0,-\sqrt{3},\sqrt{6})$, \\
    & $z=-2\sqrt{6}/3$ & $\mathbf{r}^{\text{hcp}}_{10}=\frac{a}{6}(3,\sqrt{3},-2\sqrt{6})$, $\mathbf{r}^{\text{hcp}}_{11}=\frac{a}{6}(-3,\sqrt{3},-2\sqrt{6})$, $\mathbf{r}^{\text{hcp}}_{12}=\frac{a}{3}(0,-\sqrt{3},-\sqrt{6})$\\
    \hline
    \multirow{6}{*}{dhcp}& \multirow{2}{*}{$z=0$} & $\mathbf{r}^{\text{dhcp}}_{1}=a(1,0,0)$, $\mathbf{r}^{\text{dhcp}}_{2}=\frac{a}{2}(1,\sqrt{3},0)$, $\mathbf{r}^{\text{dhcp}}_{3}=\frac{a}{2}(-1,\sqrt{3},0)$, \\
    & & $\mathbf{r}^{\text{dhcp}}_{4}=a(-1,0,0)$, $\mathbf{r}^{\text{dhcp}}_{5}=\frac{a}{2}(-1,-\sqrt{3},0)$, $\mathbf{r}^{\text{dhcp}}_{6}=\frac{a}{2}(1,-\sqrt{3},0)$, \\
    & $z=\sqrt{6}/6$ & $\mathbf{r}^{\text{dhcp}}_{7}=\frac{a}{6}(3,\sqrt{3},\sqrt{6})$, $\mathbf{r}^{\text{dhcp}}_{8}=\frac{a}{6}(-3,\sqrt{3},\sqrt{6})$, $\mathbf{r}^{\text{dhcp}}_{9}=\frac{a}{3}(0,-\sqrt{3},\frac{\sqrt{6}}{2})$, \\
    & $z=-\sqrt{6}/6$ & $\mathbf{r}^{\text{dhcp}}_{10}=\frac{a}{6}(3,\sqrt{3},-\sqrt{6})$, $\mathbf{r}^{\text{dhcp}}_{11}=\frac{a}{6}(-3,\sqrt{3},-\sqrt{6})$, $\mathbf{r}^{\text{dhcp}}_{12}=\frac{a}{3}(0,-\sqrt{3},-\frac{\sqrt{6}}{2})$, \\
    & $z=\sqrt{6}/3$ & $\mathbf{r}^{\text{dhcp}}_{13}=\frac{a}{6}(3,-\sqrt{3},2\sqrt{6})$, $\mathbf{r}^{\text{dhcp}}_{14}=\frac{a}{6}(-3,-\sqrt{3},2\sqrt{6})$, $\mathbf{r}^{\text{dhcp}}_{15}=\frac{a}{3}(0,\sqrt{3},\sqrt{6})$, \\
    & $z=-\sqrt{6}/3$ & $\mathbf{r}^{\text{dhcp}}_{16}=\frac{a}{6}(3,-\sqrt{3},-2\sqrt{6})$, $\mathbf{r}^{\text{dhcp}}_{17}=\frac{a}{6}(-3,-\sqrt{3},-2\sqrt{6})$, $\mathbf{r}^{\text{dhcp}}_{18}=\frac{a}{3}(0,\sqrt{3},-\sqrt{6})$
\end{tabular}
\end{ruledtabular}
\vspace{0.5em}
(b) Potentials
\vspace{0.5em}
\begin{ruledtabular}
\begin{tabular}{c c c}
Crystal & Atoms per Unit Cell & Potential \\
\hline
hcp & 6 & $\displaystyle{V_{\text{hcp}}(\mathbf{R})=V(\mathbf{R})+\frac{1}{3}\sum_{i=1}^6  V(\mathbf{R}+\mathbf{r}_i^{\text{hcp}})+\frac{1}{2}\sum_{i=7}^{12}  V(\mathbf{R}+\mathbf{r}_i^{\text{hcp}})}$ \\
dhcp & 12 & $\displaystyle{V_{\text{dhcp}}(\mathbf{R})=V(\mathbf{R})+\frac{1}{3}\sum_{i=1}^6  V(\mathbf{R}+\mathbf{r}_i^{\text{dhcp}})+\sum_{i=7}^{12}  V(\mathbf{R}+\mathbf{r}_i^{\text{dhcp}})+\frac{1}{2}\sum_{i=13}^{18}  V(\mathbf{R}+\mathbf{r}_i^{\text{dhcp}})}$
\end{tabular}
\end{ruledtabular}
\caption{\label{tbl:hexvecs}(a) Displacement vectors for sites in a unit cell, and (b) corresponding unit cell potentials, for the hcp and dhcp crystal structures. For the displacement vectors, the site at the origin is omitted and all vectors are given in terms of the lattice constant, $a$.}
\end{table}

\section{Details of the Density Tight-binding Model}
\label{sec:dtbm}

In this section, we outline the details of the density tight-binding configuration. As in Sec.~\ref{sec:nm}, we use simplified notation for clarity. In our model, we have a crystal of ions with tightly-bound electrons at each site. We consider each atom to be composed of a pseudopotential, which takes into account the potential of the nucleus screened by the inner electrons, and one outermost electron. As before, we employ the classical Ewald method to incorporate the long-range contribution.

\subsection{Definitions}
\label{subsec:definitions}

\subsubsection{Wavefunction}
\label{subsubsec:wavefunction}

\begin{figure}
\begin{minipage}[b]{\linewidth}
\begin{minipage}[b]{.5\linewidth}
\begin{tikzpicture}
\node at (0,0) {\centering\includegraphicsgood[width=.95\linewidth]{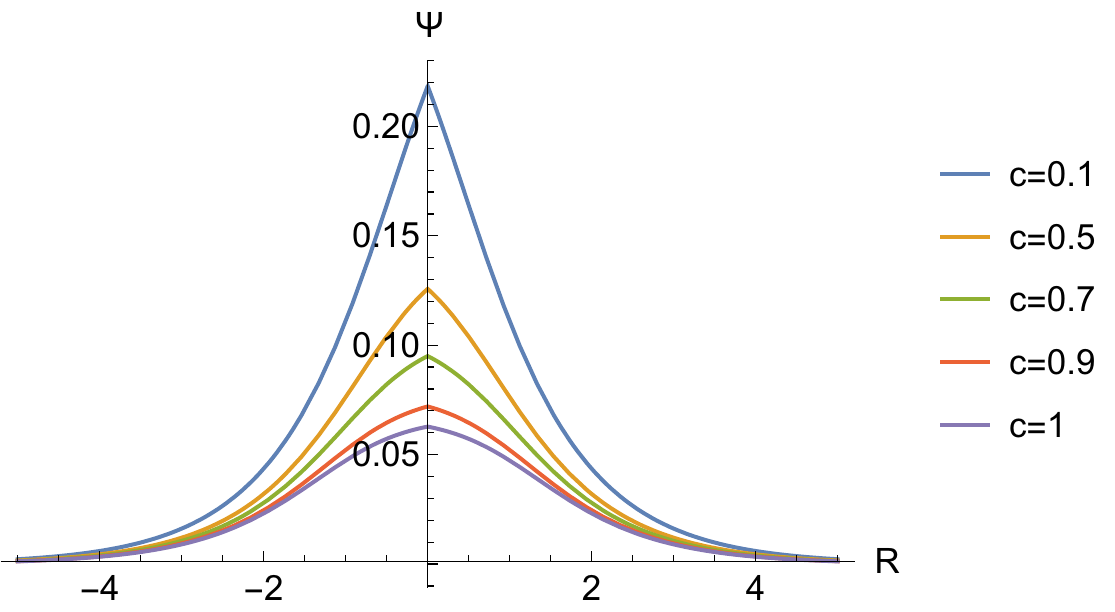}};
\node[overlay] at (-4.2,2.7) {(a)};
\label{fig:psi_c}
\end{tikzpicture}
\end{minipage}%
\begin{minipage}[b]{.5\linewidth}
\begin{tikzpicture}
\node at (0,0) {\centering\includegraphicsgood[width=.95\linewidth]{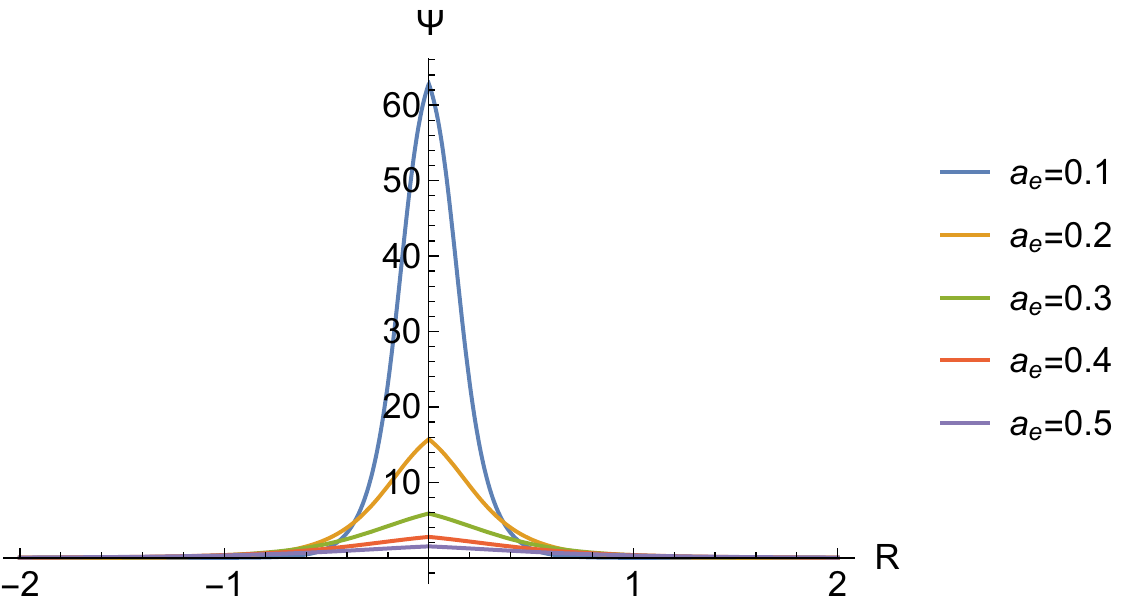}};
\node[overlay] at (-4.2,2.7) {(b)};
\label{fig:psi_ae}
\end{tikzpicture}
\end{minipage}%
\end{minipage}
\caption{\label{fig:psi}Plots of the normalized wavefunction of the valence electron under the pseudopotential of the ion, $\Psi$. The behavior of the wavefunction is shown as we (a) vary $c$ with $a_\text{e}=1$, and (b) vary $a_\text{e}$ with $c=0.1$.}
\end{figure}

We start by taking a simplified ansatz for the wavefunction of the valence electron orbital under the potential of the ion:
\begin{equation*}
\Psi(\mathbf{R};c,a_\text{e})=A\sqrt{\frac{1}{1+\exp\left(\frac{2(|\mathbf{R}|-c)}{a_\text{e}}\right)}},
\end{equation*}
where $A$ is a normalization constant, $a_\text{e}\ll R_I$ is the width of the valence electron cloud, and $0\leq c<a_\text{e}$ is the width of the core electron cloud. We choose this ansatz so that the electron density is analytically well behaved in subsequent calculations, and that in the limit of vanishing radius and large distances we recover the wavefunction of a particle in a Dirac delta potential well:
\begin{equation}
\label{eq:tbapprox}
\lim_{\substack{c\ll a_\text{e} \ll |\mathbf{R}|}}\Psi\propto e^{-|\mathbf{R}|/a_\text{e}}.
\end{equation}
This is the limit around which we will expand in the following sections. Plots of this wavefunction are shown in Fig.~\ref{fig:psi}. Since it is not possible to analytically derive an expression for the normalized wavefunction, we expand the probability density, $|\Psi|^2$ up to first order in the small parameter $(c/a_\text{e})$ and then solve the normalization condition $\int_{-\infty}^{\infty} |\Psi|^2 \mathrm{d}\mathbf{R}=a_0/a_\text{e}$, where $a_0$ is the Bohr radius. This yields a normalization constant
\begin{equation*}
A(c,a_\text{e})=\frac{2\sqrt{a_0}\left(9a_\text{e}\zeta(3)-c\pi^2\right)}{9a_\text{e}^3\sqrt{3\pi}{\zeta(3)}^{3/2}}+O\left[\left(\frac{c}{a_\text{e}}\right)^2\right],
\end{equation*}
where $\zeta(3)$ is Ap{\'e}ry's constant.

\subsubsection{Electron cloud potential and density}

The valence electron cloud (which we denote using a capital `E') has a potential given by the Coulomb potential of the single electron, $V_\text{e}(\mathbf{R})=|\mathbf{R}|^{-1}$, integrated over the density distribution of the electron cloud:
\begin{equation*}
V_\text{E}(\mathbf{R};c,a_\text{e})=\int V_\text{e}(\mathbf{R}+\mathbf{r}_\text{e}) \rho_\text{E}(\mathbf{r}_\text{e};c,a_\text{e})\,\mathrm{d}\mathbf{r}_\text{e},
\end{equation*}
where we calculate the density of the electron cloud using the normalized wavefunction defined in Sec.~\ref{subsubsec:wavefunction}:
\begin{equation*}
\rho_\text{E}(\mathbf{r}_\text{e};c,a_\text{e})=|\Psi(\mathbf{r}_\text{e};c,a_\text{e})|^2.
\end{equation*}

\subsubsection{Ion potential and density}

The ion potential is obtained by solving the time-independent Schr{\"o}dinger equation and subtracting the energy constant, such that
\begin{equation*}
V_\text{i}(\mathbf{R};c,a_\text{e})=-a_0 \left( \frac{1}{\Psi}\frac{\nabla^2}{2}\Psi - \lim_{|\mathbf{R}|\to\infty}\left( \frac{1}{\Psi} \frac{\nabla^2}{2}\Psi \right)\right).
\end{equation*}
The ion density is then subsequently obtained from Poisson's equation:
\begin{equation*}
\rho_\text{i}(\mathbf{r}_\text{i};c,a_\text{e})=-\frac{\nabla^2 V_\text{i}(\mathbf{r}_\text{i};c,a_\text{e})}{4\pi}.
\end{equation*}
Note that due to the norm conserving property of our wavefunction ansatz, the ion density satisfies the normalization condition $\int_{0}^\infty  \rho_\text{i}(r_\text{i},c,a_\text{e})4\pi r_\text{i}^2\mathrm{d} r_\text{i} = a_0/a_\text{e}$ up to first order in $(c/a_\text{e})$.

\begin{figure}
\begin{minipage}[b]{\linewidth}
\begin{minipage}[b]{.2\linewidth}
\begin{tikzpicture}
\node at (0,0) {\centering\includegraphicsgood[height=5cm]{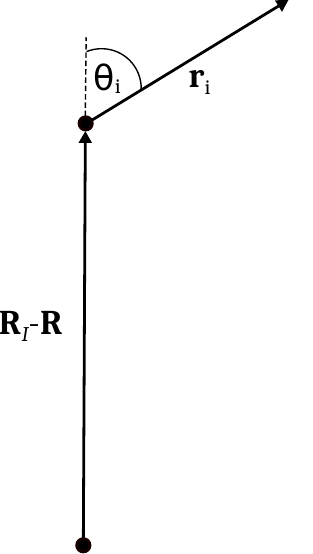}};
\node[overlay] at (-1.5,2.7) {(a)};
\label{fig:ii_contribution}
\end{tikzpicture}
\end{minipage}%
\begin{minipage}[b]{.6\linewidth}
\begin{tikzpicture}
\node at (0,0) {\centering\includegraphicsgood[height=5cm]{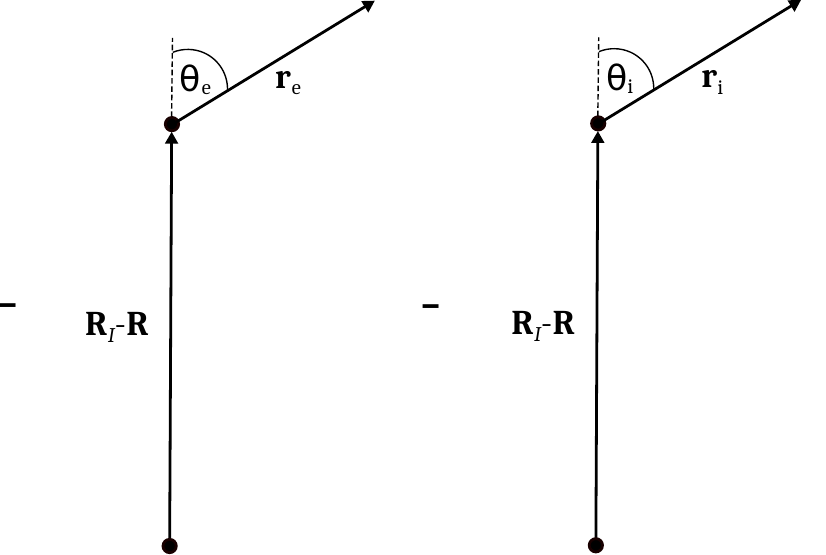}};
\node[overlay] at (-3.7,2.7) {(b)};
\label{fig:ei_contribution}
\end{tikzpicture}
\end{minipage}%
\begin{minipage}[b]{.2\linewidth}
\begin{tikzpicture}
\node at (0,0) {\centering\includegraphicsgood[height=5cm]{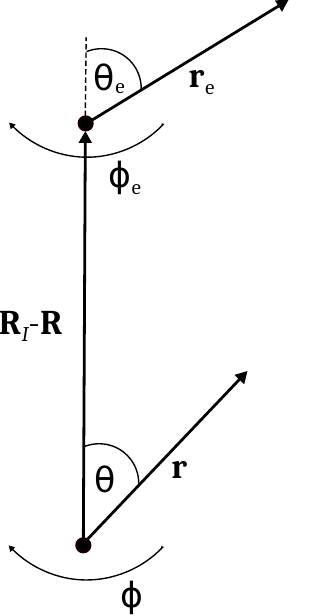}};
\node[overlay] at (-1.5,2.7) {(c)};
\label{fig:ee_contribution}
\end{tikzpicture}
\end{minipage}%
\end{minipage}
\caption{\label{fig:diagrams}Diagrams corresponding to the (a) ion-ion, (b) electron-ion, and (c) electron-electron contribution calculations. The displacement vector between ions, $\mathbf{R}-\mathbf{R}_I$, is oriented along the north pole, and the polar and azimuthal angles are defined in the range $0\leq\theta<\pi$ and $0\leq\phi<2\pi$, respectively.}
\end{figure}

\subsection{Ion-ion contribution}
\label{subsec:iicont}

First, we calculate the repulsive potential felt by an ion at position $\mathbf{R}_I$ due to an ion being displaced from the origin to a position $\mathbf{R}$. An illustration of the set-up is shown in Fig.~\ref{fig:diagrams}a. Note that we orient the displacement vector between the two ions along the north pole to simplify the calculations. In order to calculate the ion-ion potential for the whole system we then sum over all distinct atoms, such that
\begin{equation*}
E_\text{i-i}^\text{BO}(\mathbf{R};c,a_\text{e})=\sum_I \int V_\text{i}(\mathbf{R}_I-\mathbf{R}+\mathbf{r}_\text{i};c,a_\text{e}) \rho_\text{i}(\mathbf{r}_\text{i};c,a_\text{e})\,\mathrm{d}\mathbf{r}_\text{i}.
\end{equation*}
Rewriting the ion potential in terms of the scalar variables defined in Fig.~\ref{fig:diagrams}a, such that $V_\text{i}(|\mathbf{R}_I-\mathbf{R}|,\{r_\text{i},\theta_\text{i}\};c,a_\text{e})$, we may Taylor expand the ion potential up to leading order in $(r_\text{i}/|\mathbf{R}_I-\mathbf{R}|)$:
\begin{equation*}
\begin{split}
E_\text{i-i}^\text{BO}(\mathbf{R};c,a_\text{e})=&\sum_I V_\text{i}(|\mathbf{R}_I-\mathbf{R}|;c,a_\text{e})\underbrace{\int \rho_\text{i}(\mathbf{r}_\text{i};c,a_\text{e})\,\mathrm{d}\mathbf{r}_\text{i}}_{a_0/a}\\
&+2\pi \sum_I \int_{r_\text{i}=0}^{\infty}\int_{\theta_\text{i}=0}^{\pi}\left(\frac{\partial^2 V_\text{i}}{\partial r_\text{i}^2}\right)r_\text{i}^4 \rho_\text{i}(r_\text{i};c,a_\text{e})\sin(\theta_\text{i})\,\mathrm{d}\theta_\text{i}\mathrm{d}r_\text{i}+O\left[\left(\frac{r_\text{i}}{|\mathbf{R}_I-\mathbf{R}|}\right)^3\right].
\end{split}
\end{equation*}
Note that the first-order term in the expansion vanishes by symmetry. Hence the final expression for the ion-ion contribution is derived accurate to first order in $(c/a_\text{e})$ and second order in $(r_\text{i}/|\mathbf{R}_I-\mathbf{R}|)$. Taken together, this forms the leading-order analytical expansion about the density tight-binding limit introduced in Sec.~\ref{subsubsec:wavefunction}. 

\subsection{Electron-ion contribution}
\label{subsec:eicont}

The next contribution is that due to the electron-ion interaction. There are attractive potentials felt by the electron cloud due to the ions, as well as those felt by the ion due to the electron clouds. A sketch of this scenario is shown in Fig.~\ref{fig:diagrams}b, where the minus signs indicate that this is an attractive interaction. As in the previous section, we set up the general form of the electron-ion contribution as
\begin{equation*}
E_\text{e-i}^\text{BO}(\mathbf{R};c,a_\text{e})=-\sum_{I} \int V_\text{i}(\mathbf{R}_I-\mathbf{R}+\mathbf{r}_\text{e};c,a_\text{e}) \rho_\text{E}(\mathbf{r}_\text{e};c,a_\text{e})\,\mathrm{d}\mathbf{r}_\text{e}
-\sum_{I} \int V_\text{E}(\mathbf{R}_I-\mathbf{R}+\mathbf{r}_\text{i};c,a_\text{e})  \rho_\text{i}(\mathbf{r}_\text{i};c,a_\text{e})\,\mathrm{d} \mathbf{r}_\text{i}.
\end{equation*} 
It can be shown, either by symmetry or integration by parts, that this expression reduces to
\begin{equation*}
E_\text{e-i}^\text{BO}(\mathbf{R};c,a_\text{e})=-2\sum_{I} \int V_\text{i}(\mathbf{R}_I-\mathbf{R}+\mathbf{r}_\text{e};c,a_\text{e}) \rho_\text{E}(\mathbf{r}_\text{e};c,a_\text{e})\,\mathrm{d}\mathbf{r}_\text{e}.
\end{equation*}
Rewriting the ion potential in terms of scalar variables, as before, we may Taylor expand up to leading order in $(r_\text{e}/|\mathbf{R}_I-\mathbf{R}|)$:
\begin{equation*}
\begin{split}
E_\text{e-i}^\text{BO}(\mathbf{R};c,a_\text{e})=&-2 \sum_I V_\text{i}(|\mathbf{R}_I-\mathbf{R}|;c,a_\text{e})\underbrace{\int \rho_\text{E}(\mathbf{r}_\text{e};c,a_\text{e})\,\mathrm{d}\mathbf{r}_\text{e}}_{a_0/a}\\
&-4\pi \sum_I \int_{r_\text{e}=0}^{\infty}\int_{\theta_\text{e}=0}^{\pi}\left(\frac{\partial^2 V_\text{i}}{\partial r_\text{e}^2}\right)r_\text{e}^4 \rho_\text{E}(r_\text{e};c,a_\text{e})\sin(\theta_\text{e})\,\mathrm{d}\theta_\text{e}\mathrm{d}r_\text{e}+O\left[\left(\frac{r_\text{e}}{|\mathbf{R}_I-\mathbf{R}|}\right)^3\right].
\end{split}
\end{equation*}
Analogously to before, the electron-ion contribution is derived to first order in $(c/a_\text{e})$ and second order in $(r_\text{e}/|\mathbf{R}_I-\mathbf{R}|)$, which is the leading-order analytical expansion about the density tight-binding limit in this model.

\subsection{Electron-electron contribution}
\label{subsec:eecont}

Finally, we compute the repulsive electron-electron contribution to the potential. Again the displacement vector between the ions is aligned along the north pole. The valence electrons are parameterized in spherical polar coordinates around each atom, as depicted in Fig.~\ref{fig:diagrams}c. The electron-electron contribution in this case may be written as 
\begin{equation*}
E_\text{e-e}^\text{BO}(\mathbf{R};c,a_\text{e})=\sum_{I} \iint V_\text{e}(\mathbf{R}_I-\mathbf{R}+\mathbf{r}_\text{e}-\mathbf{r}) \rho_\text{E}(\mathbf{r}_\text{e};c,a_\text{e}) \rho_\text{E}(\mathbf{r};c,a_\text{e}) \,\mathrm{d}\mathbf{r}_\text{e} \mathrm{d}\mathbf{r}.
\end{equation*}
Due to the spherical symmetry of each electron cloud, this contribution reduces exactly to Coulomb repulsion, such that
\begin{equation*}
E_\text{e-e}^\text{BO}(\mathbf{R};a_\text{e})= \frac{a_0^2}{a_\text{e}^2} \sum_{I} \frac{1}{|\mathbf{R}_I-\mathbf{R}|}.
\end{equation*}
Note the total potential energy of the system at this stage, $E_\text{i-i}^\text{BO}+E_\text{e-i}^\text{BO}+E_\text{e-e}^\text{BO}$, tends to zero as $(c/a_\text{e})\to 0$ and $|\mathbf{R}|\gg a_\text{e}$. In this limit, the electrons are effectively on top of the ions and the whole system is neutral due to Gauss' theorem. 

\subsection{Pauli repulsion} 
\label{subsec:pauli}

To complement our result for the energy, we estimate the Pauli repulsion felt by the overlapping electron clouds. Since we only consider spherically symmetric (i.e.~s-type) orbitals in the toy model, this reduces to a one-dimensional problem. We consider a Dirac delta potential well, of depth $g$, inside an infinite square well, such that:
\begin{equation*}
V_\text{well}(x)=\begin{cases}
-g\delta(x), & |x|=0, \\
0, & 0<|x|<L, \\
\infty, & |x|\geq L.
\end{cases}
\end{equation*}
In this scenario, $g$ determines how tightly bound the electrons are to their respective atoms, and $L$ represents the effective radius for the electron clouds. As $L$ is reduced, the bound state energy is increased -- this represents the energy increase due to the Pauli repulsion of overlapping orbitals. 

The wavefunction takes the form $\Psi\propto \sinh (k(L-|x|))$ inside the infinite well, where $k$ is the wave number. Considering the derivative continuity of the wavefunction at the origin, we derive the transcendental equation $\tanh y = \chi y$, where we have defined $y\equiv k L$ and $\chi \equiv \hbar^2 / m g L$. We can derive an analytical form for the solution, and hence the scaling behavior of the energy with $L$, by finding the lowest root with a Newton-Raphson scheme. The iterative equation for the root is then 
\begin{equation*}
y_{n+1}=y_n-\frac{\tanh y_n - y_n\chi}{\text{sech}^2 y_n - \chi},
\end{equation*}
where $n\in\mathbb{Z}^+$. Since we are interested in the regime where the wavefunction is significantly influenced by the boundary wall, we take $y_n$ to be small. Additionally, we are interested in the limit when Pauli repulsion is dominant i.e.~when $L$ is small. Taking these limits together, we find that $y_\infty = \sqrt{3\chi / 2}$. Hence the energy of the bound state is
\begin{equation*}
E_\text{Pauli}^\text{BO}=\frac{\hbar^2}{2 m L^2}y_\infty^2=\frac{3 \hbar^2}{4 m^{3/2}\sqrt{2 E_\text{b}}L^3},
\end{equation*}
where we define the binding energy of an isolated Dirac delta potential well as $E_\text{b}\equiv mg^2 / 2 \hbar^2$. In the density tight-binding approximation, the wavefunction takes the form $\Psi\propto \exp (-m^{1/2}\sqrt{2E_\text{b}}L/\hbar)$. Comparing this to the form of the wavefunction in Eq.~\ref{eq:tbapprox} allows us to make the identification $E_\text{b}\sim a_\text{e}^{-2}$ up to physical constants. Hence, in atomic units, the energy gain due to Pauli repulsion becomes
\begin{equation*}
E_\text{Pauli}^\text{BO}=\frac{3 a_\text{e}}{4 L^3}.
\end{equation*}
Note that due to the differences in unit cell geometry, the lattice constant cannot be directly compared between the various crystal structures. For this, we may examine the optimal effective radius of each atom in a spherical packing, defined as
\begin{equation*}
r_\text{eff}=\left(\frac{3}{4\pi}\frac{V_\text{u}}{N_\text{u}}\right)^{1/3},
\end{equation*}
where $V_\text{u}$ is the optimal volume of the unit cell, and $N_\text{u}$ is the number of atoms enclosed.  In place of $L$, we evaluate $E_\text{Pauli}^\text{BO}$ at the effective optimum radius. This rudimentary approximation for the Pauli repulsion allows us to analytically capture the scaling behavior as the lattice constant is reduced.

\subsection{Crystal relaxation}
\label{subsec:relax}

Let us define the BO energy of the system as
\begin{equation*}
E^\text{BO}(\mathbf{R};c,a_\text{e})=E_\text{i-i}^\text{BO}(\mathbf{R};c,a_\text{e})+E_\text{e-i}^\text{BO}(\mathbf{R};c,a_\text{e})+E_\text{e-e}^\text{BO}(\mathbf{R};a_\text{e})+E_\text{Pauli}^\text{BO}(a,a_\text{e}).
\end{equation*}
Note that there is an implicit lattice constant dependence in the first three terms in the form of the potentials, as well as in the lattice summations. Once we have calculated an analytical form for the BO energy of the system as a function of the displacement of the central atom, $\mathbf{R}$, and implicitly the lattice constant, $a$, we then compute the optimal lattice constant such that:
\begin{equation*}
a_\text{min}=\underset{a\in(a_\text{e},\infty)}{\text{argmin}}\left( E^\text{BO} \right).
\end{equation*}
We subsequently relax the system to this lattice constant, compute the nuclear kinetic energy, and evaluate the total energy at a given $\mathbf{R}$. This renders the total energy, $E$, as a function of $c$ and $a_\text{e}$ only. 

\section{Oscillatory Electron Density in the Density Nearly-free Electron Model} 
\label{sec:edanfem}

\begin{table}
\begin{ruledtabular}
\begin{tabular}{c c c}
Crystal & $\rho_\text{E}^\text{osc}(\mathbf{r})/u$ \\
\hline
cub/bcc/fcc & $\displaystyle{\frac{1}{\tilde{N}_\text{c.s.}}\sum_{i=1}^{\tilde{N}_\text{c.s.}}\cos(\mathbf{r}\cdot\tilde{\mathbf{r}}^\text{c.s.}_i)}$ \\
dia & $\displaystyle{\frac{1}{8}\left[\sum_{i=1}^{8}\cos(\mathbf{r}\cdot\tilde{\mathbf{r}}^\text{dia}_i)+\sum_{i=1}^{8}\cos\left(\left(\mathbf{r}-\mathbf{r}^\text{dia}_{13}
\right)\cdot\tilde{\mathbf{r}}^\text{dia}_i\right)\right]}$ \\
hcp & $\displaystyle{\frac{A_\text{hcp}}{6}\left[\sum_{i=1}^{6}\cos(\mathbf{r}\cdot\tilde{\mathbf{r}}^\text{hcp}_i)\cos\left( \frac{3\pi}{\sqrt{6}a}z\right)+\sum_{i=1}^{6}\cos\left(\left(\mathbf{r}-\mathbf{r}^\text{hcp}_7
\right)\cdot\tilde{\mathbf{r}}^\text{hcp}_i\right)\cos\left( \frac{3\pi}{\sqrt{6}a}\left(z-\frac{\sqrt{6}a}{3}\right)\right)\right]}$ \\
dhcp & $\begin{aligned} \displaystyle{
\frac{1}{6}\left[\frac{1}{3}\sum_{i=1}^{6}\cos(\mathbf{r}\cdot\tilde{\mathbf{r}}^\text{dhcp}_i)\cos\left( \frac{3\pi}{\sqrt{6}a}z\right)\right.} & \displaystyle{\left.+\sum_{i=1}^{6}\cos\left(\left(\mathbf{r}-\mathbf{r}^\text{dhcp}_7
\right)\cdot\tilde{\mathbf{r}}^\text{dhcp}_i\right)\cos\left( \frac{\sqrt{6}\pi}{a}\left(z-\frac{a}{\sqrt{6}}\right)\right)\right.} \\
& \displaystyle{\left.+\frac{1}{3}\sum_{i=1}^{6}\cos\left(\left(\mathbf{r}-\mathbf{r}^\text{dhcp}_{14}
\right)\cdot\tilde{\mathbf{r}}^\text{dhcp}_i\right)\cos\left( \frac{3\pi}{\sqrt{6}a}\left(z-\frac{\sqrt{6}a}{3}\right)\right)\right]} \end{aligned}$
\end{tabular}
\end{ruledtabular}
\caption{\label{tbl:osc}Oscillatory part of the electron cloud density in the density nearly-free electron model, $\rho_\text{E}^\text{osc}$, in units of the oscillation strength, $u$. $\tilde{N}_\text{c.s.}$ is the number of displacement vectors, and $\{\tilde{\mathbf{r}}^\text{c.s.}_i\}$ the set of displacements, in a unit cell of the reciprocal lattice. The vectors $\mathbf{r}^\text{c.s.}_i$ are defined in Tables~\ref{tbl:cubicvecs} and~\ref{tbl:hexvecs}. The normalization constant, $A_\text{hcp}=2/3$, is chosen such that $\max(\rho_\text{E}^\text{osc})=u$ for all crystal structures.}
\end{table}

In order to approximate the oscillatory part of the electron cloud density in the density nearly-free electron model, we consider Fourier transforms of the reciprocal lattices, as shown in Table~\ref{tbl:osc}. For crystals with a single-ion basis, the resulting function has a simple form. However, for crystals with more than one ion in the basis, we consider a superposition of multiple offset lattices\footnote{Note that the offsets given in Table~\ref{tbl:osc} are not unique.}; with modulation along the z-axis, where appropriate. The functions, $\rho_\text{E}^\text{osc}$, are scaled such that $\max(\rho_\text{E}^\text{osc})=u$ for all crystal structures. Over a unit cell, all of the functions integrate to zero.

\bibliographystyle{apsrev4-1}
\bibliography{ms}